\documentclass{egpubl}
\usepackage{egsr2021}
\usepackage{amsfonts}
 
\JournalSubmission    
\usepackage[T1]{fontenc}
\usepackage{dfadobe}

\biberVersion
\BibtexOrBiblatex
\usepackage[backend=biber,bibstyle=EG,citestyle=alphabetic,backref=true]{biblatex} 
\electronicVersion
\PrintedOrElectronic

\ifpdf \usepackage[pdftex]{graphicx} \pdfcompresslevel=9
\else \usepackage[dvips]{graphicx} \fi
\usepackage{xcolor}

\usepackage{caption}
\usepackage{subcaption}
\usepackage{multirow}
\usepackage{booktabs}
\usepackage{amsfonts}

\usepackage{egweblnk} 

\title[Dynamic Diffuse Global Illumination Resampling]%
      {Dynamic Diffuse Global Illumination Resampling}

\author[Z. Majercik \& T. M\"uller \& A. Keller \& D. Nowrouzezahrai \& M. McGuire]
{\parbox{\textwidth}{\vspace*{-1.6cm}\centering Zander Majercik$^{1}$, Thomas M\"uller$^{1}$, Alexander Keller$^{1}$, Derek Nowrouzezahrai$^{2}$, and Morgan McGuire$^{3}$
        }
        \\
{\parbox{\textwidth}{\vspace*{-1.8cm}\centering $^1$NVIDIA
         $^2$McGill University
         $^3$Roblox
       }
}
}

\gdef\cropInsets{0}

\usepackage{tabularx}
\newcolumntype{Y}{>{\centering\arraybackslash}X}

\usepackage{xargs}
\usepackage{mathtools}
\usepackage{amsmath}
\usepackage{microtype}
\usepackage{units}
\usepackage[binary-units]{siunitx}
\usepackage{overpic}
\usepackage{cancel}
\usepackage{wrapfig}
\usepackage{placeins}
\usepackage{empheq}
\usepackage{xspace}

\usepackage{url} 
\usepackage[utf8]{inputenc}

\def\commentType{1}

\ifnum\commentType=0
    \newcommandx{\customComment}[3]{}
    \newcommandx{\customTODO}[3]{}
\fi
\ifnum\commentType=1
    \newcommandx{\customComment}[3]{\textcolor{#2}{\textsl{#1: #3}}}
    \newcommandx{\customTODO}[3]{\textcolor{#2}{\textsl{#1: #3}}}
\fi
\ifnum\commentType=2
    \usepackage{pdfcomment}
    \newcommandx{\customComment}[3]{\pdfcomment[icon=Comment,opacity=0.5,color=#2,author=#1]{#3}}
    \newcommandx{\customTODO}[3]{\pdfcomment[icon=Note,opacity=0.5,color=#2,author=#1]{#3}}
\fi
\ifnum\commentType=3
    \usepackage{pdfcomment} 
    \usepackage{todonotes} 
    \usepackage{silence}
    \newcommandx{\customComment}[3]{\todo[color=#2!40,size=\small]{\textbf{#1:} #3}}
    \newcommandx{\customTODO}[3]{\todo[color=#2!40,size=\small]{\textbf{#1:} #3}}
\fi

\let\originalleft\left 
\let\originalright\right 
\renewcommand{\left}{\mathopen{}\mathclose\bgroup\originalleft} 
\renewcommand{\right}{\aftergroup\egroup\originalright} 

\definecolor{amber}{rgb}{1.0, 0.49, 0.0}
\definecolor{darkgreen}{rgb}{0.0, 0.5, 0.0}
\newcommandx{\All}[1]{\customComment{All}{red}{#1}}
\newcommandx{\Jan}[1]{\customComment{Jan}{magenta}{#1}}
\newcommandx{\Alex}[1]{\customComment{Alex}{blue}{#1}}
\newcommandx{\Fabrice}[1]{\customComment{Fabrice}{amber}{#1}}
\newcommandx{\Thomas}[1]{\customComment{Thomas}{darkgreen}{#1}}

\newcommandx{\TODO}[1]{\customTODO{TODO}{red}{#1}}
\newcommandx{\JanTODO}[1]{\customTODO{Jan}{magenta}{#1}}
\newcommandx{\FabriceTODO}[1]{\customTODO{Fabrice}{amber}{#1}}
\newcommandx{\ThomasTODO}[1]{\customTODO{Thomas}{darkgreen}{#1}}

\newcommand{\IGNORE}[1]{}

\newcommand{\REMOVE}[1]{} 

\def\equationautorefname~#1\null{%
  Equation~(#1)\null
}

\newcommand{\Diff}[1]{\,\mathrm{d}#1}

\newcommand{\R}{\mathbb{R}}

\newcommand{\pos}{\mathbf{x}}
\newcommand{\posy}{\mathbf{y}}
\newcommand{\posz}{\mathbf{z}}
\newcommand{\diro}{\omega_\mathrm{o}}
\newcommand{\dir}{\omega}
\newcommand{\diri}{\omega_\mathrm{i}}
\newcommand{\dirr}{\omega_\mathrm{r}}

\newcommand{\normal}{\mathbf{n}}

\newcommand{\Lddgi}{L_\text{DDGI}}
\newcommand{\LddgiP}{L_\text{DDGI+}}
\newcommand{\Eddgi}{E_\text{DDGI}}
\newcommand{\EddgiP}{E_\text{DDGI+}}

\newcommand{\bsdf}{f}
\newcommand{\bsdfDiff}{\bsdf_{\mathrm{d}}}

\newcommand{\bsdfGlossy}{\bsdf_{\mathrm{g}}}

\newcommand{\AreaHeuristic}{a}

\newcommand{\LivingRoom}{\textsc{Living Room}\xspace}
\newcommand{\PinkRoom}{\textsc{Pink Room}\xspace}

\newcommand{\RoomDoor}{\textsc{Room Door}\xspace}
\newcommand{\SplitRoom}{\textsc{Split Room}\xspace}
\newcommand{\RedBall}{\textsc{Red Ball}\xspace}
\newcommand{\GreekVilla}{\textsc{Greek Villa}\xspace}

\gdef\useCroppedImages{1}

\usepackage[export]{adjustbox}
\usepackage{calc}
\usepackage{tabularx}
\usepackage{tikz}
\usepackage{xstring}

\newlength{\beautyHeight}
\newlength{\beautyPixWidth}
\newlength{\beautyPixHeight}
\newlength{\insetvsep}
\gdef\useInsetA{0}
\gdef\useInsetB{0}
\gdef\useInsetC{0}

\newcommand{\setInset}[6]{%
    \expandafter\gdef\csname useInset#1\endcsname{1}%
    \expandafter\gdef\csname inset#1Color\endcsname{#2}%
    \expandafter\gdef\csname crop#1X\endcsname{#3}%
    \expandafter\gdef\csname crop#1Y\endcsname{#4}%
    \expandafter\gdef\csname crop#1W\endcsname{#5}%
    \expandafter\gdef\csname crop#1H\endcsname{#6}%
}

\newcommand{\unsetInset}[1]{%
    \expandafter\gdef\csname useInset#1\endcsname{0}%
}

\newcommand{\addBeautyCrop}[8]{%
    \pdfpxdimen=\dimexpr 1 in/72\relax
    \def\beauty{%
        \let\cropR\relax%
        \let\cropB\relax%
        \newlength\cropR%
        \newlength\cropB%
        \setlength\cropR{{#3 px}-{#5 px}-{#7 px}}%
        \setlength\cropB{{#4 px}-{#6 px}-{#8 px}}%
        \sbox0{\includegraphics[width=#2\textwidth,trim={#5px {\cropB} {\cropR} #6px},clip]{#1}}%
        \begin{tikzpicture}
            \node[anchor=north west,inner sep=0] at (0,0) {\usebox0};
            \begin{scope}[x=\wd0/#7, y=\ht0/#8]
            \if\useInsetA1{
                \draw[\insetAColor,thick] (\cropAX-#5,-\cropAY+#6) rectangle + (\cropAW,-\cropAH);
            }\fi
            \if\useInsetB1{
                \draw[\insetBColor,thick] (\cropBX-#5,-\cropBY+#6) rectangle + (\cropBW,-\cropBH);
            }\fi
            \if\useInsetC1{
                \draw[\insetCColor,thick] (\cropCX-#5,-\cropCY+#6) rectangle + (\cropCW,-\cropCH);
            }\fi
            \end{scope}
        \end{tikzpicture}
    }%
    \setlength\beautyHeight{\heightof{\beauty}}%
    \setlength\beautyPixWidth{#3px}%
    \setlength\beautyPixHeight{#4px}%
    \global\beautyHeight=\beautyHeight%
    \global\beautyPixWidth=\beautyPixWidth%
    \global\beautyPixHeight=\beautyPixHeight%
    \begin{adjustbox}{valign=t}
        \beauty{}
    \end{adjustbox}
}

\newcommand{\trimInset}[6]{%
    \let\cropR\relax%
    \let\cropB\relax%
    \newlength\cropR%
    \newlength\cropB%
    \setlength\cropR{{\beautyPixWidth}-{#3 px}-{#5 px}}%
    \setlength\cropB{{\beautyPixHeight}-{#4 px}-{#6 px}}%
    \color{#2}%
    \fbox{\includegraphics[width=1\linewidth,trim={{#3 px} {\cropB} {\cropR} {#4 px}},clip]{#1}}%
}

\newcommand{\addInset}[2]{%
    \color{#2}%
    \fbox{\includegraphics[width=1\linewidth]{#1}}%
}

\newcommand{\auxtimes}{x}
\newcommand{\auxplus}{+}
\newcommand{\auxspace}{ }

\newcommand{\addInsets}[1]{%
    \begin{adjustbox}{valign=t}
        \StrSubstitute{#1}{.}{-}[\baseFileName]
        \begin{adjustbox}{totalheight=1\beautyHeight,tabular={c}}
            \if\useInsetA1%
                \def\cropfile{\baseFileName-\cropAW\auxtimes\cropAH\auxplus\cropAX\auxplus\cropAY-crop}
                \if\cropInsets1
                    \immediate\write18{convert #1 -crop \cropAW\auxtimes\cropAH\auxplus\cropAX\auxplus\cropAY\auxspace -filter point -resize 800\% \cropfile.jpg}
                \fi
                \if\useCroppedImages1
                    \addInset{\cropfile.jpg}{\insetAColor}
                \else
                    \trimInset{#1}{\insetAColor}{\cropAX}{\cropAY}{\cropAW}{\cropAH}%
                \fi%
            \fi%
            \if\useInsetB1%
                \if\useInsetA1\\[\insetvsep]\fi%
                \def\cropfile{\baseFileName-\cropBW\auxtimes\cropBH\auxplus\cropBX\auxplus\cropBY-crop}
                \if\cropInsets1
                    \immediate\write18{convert #1 -crop \cropBW\auxtimes\cropBH\auxplus\cropBX\auxplus\cropBY\auxspace -filter point -resize 800\% \cropfile.jpg}
                \fi
                \if\useCroppedImages1
                    \addInset{\cropfile.jpg}{\insetBColor}
                \else
                    \trimInset{#1}{\insetBColor}{\cropBX}{\cropBY}{\cropBW}{\cropBH}%
                \fi%
            \fi%
            \if\useInsetC1%
                \if\useInsetB1\\[\insetvsep]\fi%
                \def\cropfile{\baseFileName-\cropCW\auxtimes\cropCH\auxplus\cropCX\auxplus\cropCY-crop}
                \if\cropInsets1
                    \immediate\write18{convert #1 -crop \cropCW\auxtimes\cropCH\auxplus\cropCX\auxplus\cropCY\auxspace -filter point -resize 800\% \cropfile.jpg}
                \fi
                \if\useCroppedImages1
                    \addInset{\cropfile.jpg}{\insetCColor}
                \else
                    \trimInset{#1}{\insetCColor}{\cropCX}{\cropCY}{\cropCW}{\cropCH}%
                \fi%
            \fi%
        \end{adjustbox}
    \end{adjustbox}
}

\definecolor{mathematicaBlue}{rgb}{0.38, 0.51, 0.71}
\definecolor{mathematicaOrange}{rgb}{0.88, 0.61, 0.14}
\definecolor{mathematicaGreen}{rgb}{0.56, 0.69, 0.19}
\definecolor{mathematicaRed}{rgb}{0.92,0.39, 0.21}
\definecolor{mathematicaPurple}{rgb}{0.53, 0.47, 0.7}

\definecolor{cvintegrand}{rgb}{1.0, 0.65, 0.0}
\definecolor{cvg}{rgb}{0.5, 0.0, 0.5}
\definecolor{cvG}{rgb}{0.67, 0.14, 0.19}

\definecolor{cvdifference}{rgb}{1.0, 0.65, 0.0}
\definecolor{cvpdf}{rgb}{0.5, 0.0, 0.5}

\begin{document}

\teaser{\vspace*{-0.8cm}
\setlength{\fboxrule}{1pt}%
\setlength{\tabcolsep}{1pt}%
\renewcommand{\arraystretch}{1}%
\small%
\centering%
\vspace{-10mm}
\begin{tabular}{cccccccc}%

                    &
                    
                   Ours + Denoising &
                    
                    &
                    
                    &
                    
                    \hspace{-4.5cm}Offline Reference
                \\%

    \multicolumn{5}{c}{\fcolorbox{black}{black}{\setlength{\fboxsep}{1.5pt}\begin{overpic}[width=\linewidth, trim=0 260 0 90, clip]{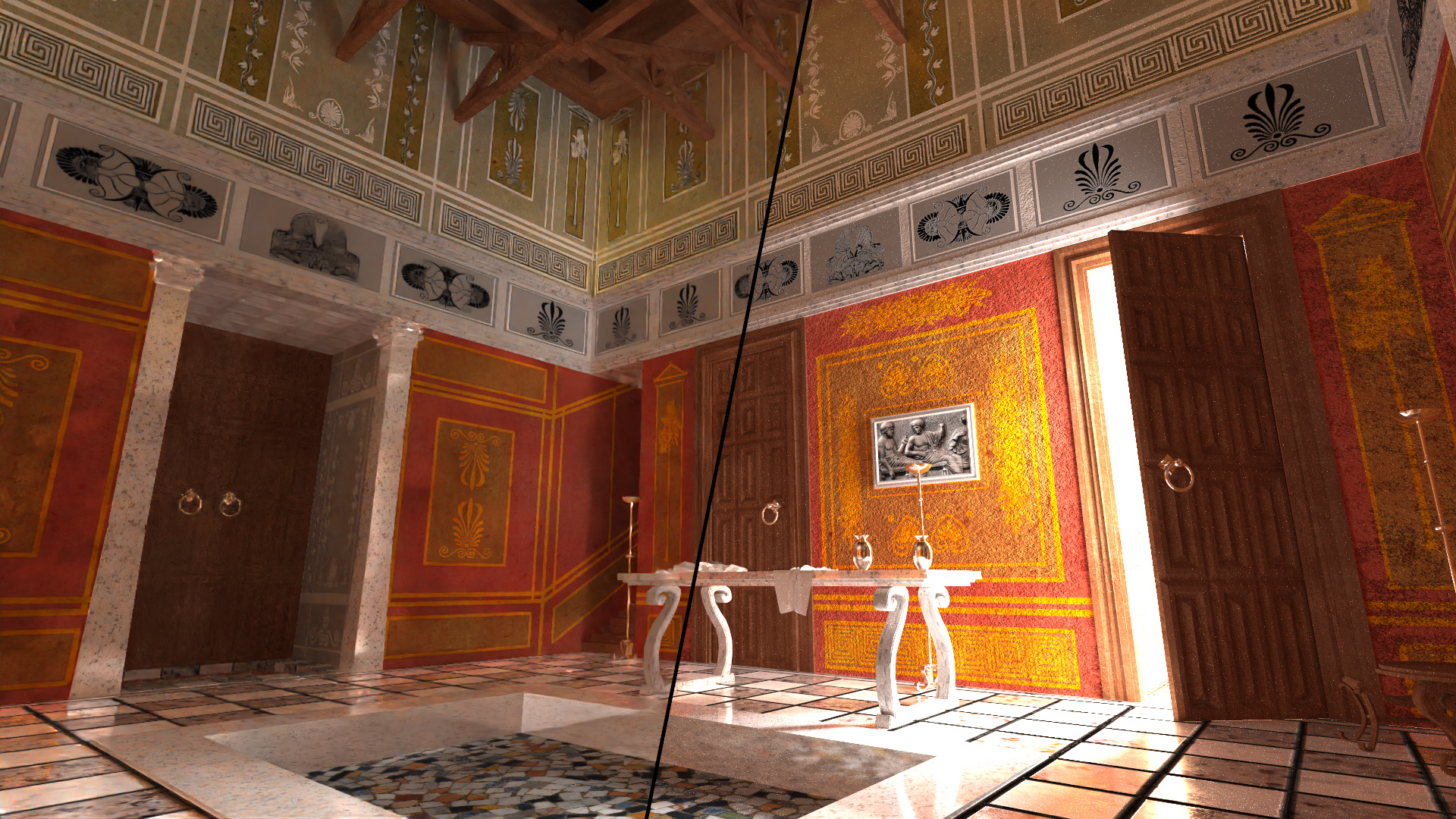}
        \put(38.54166666666667, 35.15625){\makebox(0,0){\tikz\draw[red,ultra thick] (0,0) rectangle (0.10312500000000001\linewidth, 0.041249999999999995\linewidth);}}
        \put(35.104166666666664, 22.916666666666664){\makebox(0,0){\tikz\draw[orange,ultra thick] (0,0) rectangle (0.10312500000000001\linewidth, 0.041249999999999995\linewidth);}}
    \end{overpic}}}
    \\%
    \fcolorbox{red}{red}{\includegraphics[width=0.19\linewidth]{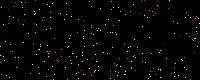}} &
    \fcolorbox{red}{red}{\includegraphics[width=0.19\linewidth]{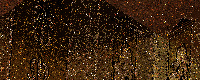}} &
    \fcolorbox{red}{red}{\includegraphics[width=0.19\linewidth]{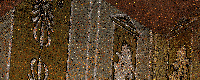}} &
    \fcolorbox{red}{red}{\includegraphics[width=0.19\linewidth]{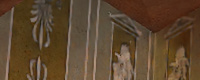}} &\hspace{0.5mm}\vline\hspace{1mm}\vline\hspace{0.5mm}
    \fcolorbox{red}{red}{\includegraphics[width=0.19\linewidth]{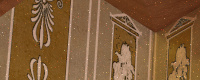}}
    \\%
    \fcolorbox{orange}{orange}{\includegraphics[width=0.19\linewidth]{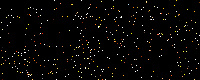}} &
    \fcolorbox{orange}{orange}{\includegraphics[width=0.19\linewidth]{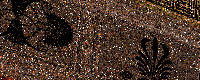}} &
    \fcolorbox{orange}{orange}{\includegraphics[width=0.19\linewidth]{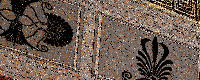}} &
    \fcolorbox{orange}{orange}{\includegraphics[width=0.19\linewidth]{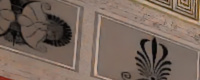}} &\hspace{0.5mm}\vline\hspace{1mm}\vline\hspace{0.5mm}
    \fcolorbox{orange}{orange}{\includegraphics[width=0.19\linewidth]{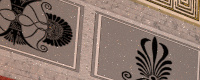}}
    \\%
            Path tracing + ReSTIR
                    &
                    + Secondary DDGI (ours)
                    &
                    + DDGI Resampling (ours)
                    &
                    + Denoising (ours)
                    & Reference
                \\%
            22.5 ms&
            12.8 ms&
            18.4 ms&
            26.6 ms&
        \\%
\end{tabular}
  \caption{The indirectly lit \GreekVilla{} scene rendered at $1920\times1080$ with 1 sample per pixel (spp) on an i7 6800k CPU and RTX 3090 GPU.
  Direct illumination resampling, such as ReSTIR~\cite{bitterli20spatiotemporal} (leftmost inset), reduces noise at the primary path vertex but does not affect the remainder of the path which is both noisy and expensive to trace.
  Therefore, we propose replacing the remainder of the path, starting from the secondary vertex, with a cheap, noise-free approximation: an extended variant of DDGI~\cite{Majercik2021ScalingGI} (second inset).
  Our key observation is that DDGI thereby acts as a light source, allowing us to include it in the resampling algorithm (third inset).
  This unifies the sampling of direct and indirect illumination.
  Combined with denoising (fourth inset), scenes with complex indirect illumination such as this one can be rendered in real time from 1 spp, with quality approaching offline path tracing with hundreds of samples per pixel (fifth inset).
  }\label{fig:teaser}
  \vspace{0.3cm}
}

\maketitle
\begin{abstract}
Interactive global illumination remains a challenge in radiometrically- and geometrically-complex scenes. Specialized sampling strategies are effective for specular and near-specular transport because the scattering has relatively low directional variance per scattering event. In contrast, the high variance from transport paths comprising multiple rough glossy or diffuse scattering events remains notoriously difficult to resolve with a small number of samples. 
We extend unidirectional path tracing to address this by combining screen-space reservoir resampling and sparse world-space probes, significantly improving sample efficiency for transport contributions that terminate on diffuse scattering events.  Our experiments demonstrate a clear improvement -- at equal time and equal quality -- over purely path traced and purely probe-based baselines. Moreover, when combined with commodity denoisers, we are able to interactively render global illumination in complex scenes.


\end{abstract}  

\section{Introduction}

Modern physically-based production renderers rely primarily on variants of the unidirectional path tracing algorithm~\cite{ToGProduction}, in which paths are traced from the camera and scattered in the scene until they reach an emitter. Path tracing scenes with high geometric and radiometric complexity can result in high-variance (i.e.,\ noisy) images unless they take a prohibitively large number of samples.
To alleviate this, many Monte Carlo variance reduction-based solutions have been proposed, such as variants of importance sampling~\cite{Veach95Multiple,Owen98Safe,Talbot2005,bitterli20spatiotemporal} and path guiding~\cite{Vorba2019PGP}.
However, even with advanced sampling techniques in place, hundreds of samples per pixel are typically required to produce a converged image, which is prohibitive for real-time applications.

Even though recent improvements in GPU ray tracing and effective denoising have greatly improved path tracing performance for complex scenes, real-time path tracing budgets remain limited to only few indirect scattering events (often just one) and scenes with only modestly complex illumination. As such, interactive graphics methods necessarily resort to approximate global illumination methods \emph{in addition} to importance sampling. One such recent approach combines world-space irradiance probes with visibility-aware interpolation to rapidly approximate multi-bounce indirect illumination (DDGI)~\cite{Majercik2019Irradiance}.
Crucially, the normal-dependent irradiance can also be interpreted as cosine-prefiltered radiance~\cite{Majercik2021ScalingGI} in the normal direction, allowing its use in both rough glossy as well as Lambertian diffuse transport.
As with most approximate solutions, these \textit{dynamic probes} introduce bias -- in the form of the transport that is computed -- in exchange for a noise-free and fast result.

In contrast to interactive indirect illumination sampling, the sample efficiency and approachable scene complexity for ray-traced direct illumination has been greatly improved by recent a spatio-temporal resampling scheme (ReSTIR)~\cite{bitterli20spatiotemporal}. 
In the graphics pipeline, sampling is followed by full-screen post-processing, including denoising~\cite{Bitterli2016Denoising,Boughida2017Denoising,Schied2017Denoising,Vogels2018Denoising,Bako2017Denoising,Chaitanya2017Denoising,Mara2017Denoise,Xu2019Denoising,Huo2021DenoisingSurvey}.
So, the goal of a modern real-time sampling algorithm is not a fully converged image, but rather one with sufficiently low noise that after post-processing it is acceptably close to a converged image for the application. We address the efficient sampling problem in this work, assuming a downstream denoiser. We note that today's commodity denoisers both desirably conceal sampling noise and also undesirably exhibit various residual temporal errors that are beyond the scope of this work. Hence, we show the impact of denoising in Fig.~\ref{fig:teaser} for a static image to validate the unconverged targets, and then show the pre-denoising output of our method in all other results.

We present a new real-time global illumination method that combines the advantages of spatio-temporal resampling (improved sample efficiency for stochastic estimators) and dynamic diffuse probe volumes (smooth, interpolation-friendly transport proxies). Our method is efficient, treats direct and multi-bounce indirect effects, and is robust to radiometrically- and geometrically-complex settings. 
Na\"\i{}vely combining DDGI with ReSTIR (see Fig.~\ref{fig:main-result1spp}, ``Primary DDGI'' column) will neither eliminate the probe grid artifacts nor address the discrepancy in their respective artifact/noise characteristics: DDGI's noise-free and biased indirect illumination, and ReSTIR's noisy and unbiased direct illumination.

We alleviate this discrepancy by postponing DDGI queries by one bounce (from the eye) -- where sample contributions are evaluated during ReSTIR
(see Fig.~\ref{fig:main-result1spp}, ``DDGI in ReSTIR'' column) -- similarly to \textit{final gathering} with photon mapping~\cite{Jensen1996PhotonMap}. This hides a large portion of DDGI's bias and allows us to combine direct and indirect samples during resampling, selecting each sample proportional to its \textit{global} transport contribution.
Where the original ReSTIR approach would normally resample direct illumination, it now treats both direct and indirect samples, each with similar noise characteristics due to the final gathering-like postponed DDGI sample, resulting in a sample efficient global shading estimate that is amenable to commodity denoising methods (see Fig.~\ref{fig:teaser}).

Our method takes a principled approach to balancing the performance-quality trade-off inherent to any combination of DDGI and ReSTIR, as evidenced in our analysis and benchmark against
purely path traced and purely probe-based
variants (Sec.~\ref{Sec:Discussion}). Concisely, our contributions are as follows:
\begin{itemize}
    \item a categorization of contributions from each constituent technique,
    \item an in-depth analysis of sampling-based methods applied to these contributions,
    \item an interactive global illumination algorithm, motivated by our analysis, that combines resampling with a global transport approximation.
\end{itemize}

\section{Background and Related Work}

\paragraph*{(Ir)radiance Caching.}
Most modern real-time global illumination caches have their roots in Ward's seminal irradiance cache~\cite{Ward88IrradianceCache} and other early approaches to precomputed light transport~\cite{Arvo86backwardray,heckbert1990adaptive}.
While initial work focused on adapting precomputed light transport to the harsh constraints of real-time rendering in \emph{static} settings~\cite{Abrash:1997:quake,Oat:2005:irradiance,GeomericsEnlighten}, many recent techniques offer at least partial dynamics, such as dynamic lighting through precomputed radiance transfer~\cite{Sloan:2002:precomputed}.
Orthogonally, radiance caching techniques overcome the Lambertian assumption by including the directional domain~\cite{Krivanek:2005:caching}.
Although today's selection of high-quality \emph{partially dynamic} techniques is vast~\cite{Gilabert:2012:deferred,Silvennoinen:2017,Seyb:2020,Vardis:2014:radiance,Scherzer:2012,Rehfeld:2014,Schwarzhaupt:2012:hessian}, our goal is \emph{fully dynamic} global illumination with as few constraints as possible.

To this end, we build upon the DDGI volume~\cite{Majercik2019Irradiance}, which consists of a 3D grid of directionally resolved irradiance probes that is updated in real-time through hardware ray-tracing.
Crucially, DDGI contains visibility information to prevent light leaking and it can approximate sufficiently rough \emph{glossy} transport by re-purposing its directional dependence~\cite{Majercik2021ScalingGI}.
These features make it -- with mild modifications (Sec.~\ref{sec:algorithm}) -- a reasonably versatile approximation of global illumination for most phenomena other than specular reflections.
By querying DDGI \emph{at secondary path vertices}, we employ a similar strategy to final gathering to conceal DDGI's limitations (e.g.,\ coarse spatial discretization) behind the blurring effect of the primary scattering interaction at the cost of sampling noise. We then address the sampling noise by applying spatio-temporal reservoir sampling~\cite{bitterli20spatiotemporal}.

M\"uller et al.~\cite{mueller2021realtime} use a real-time trained neural radiance cache to approximate fully-dynamic global illumination.
They evaluate their cache at later path vertices to hide the artifacts of their neural network, but unlike this paper, they do not employ specialized importance sampling.
Crucially, their method is compatible with ours in that the caches can be interchanged in future investigations.

\paragraph*{Importance Sampling.}
In offline rendering, importance sampling is the dominant technique for reducing sampling noise: the closer the sampling distribution matches the distribution of light, the less noise there is~\cite{VeachPhD}.
With the advent of hardware ray tracing, it is now important to adapt established techniques to the real-time setting~\cite{AreWeDoneWithRayTracing}.

Like with caching techniques, spatio-temporal reuse is key for high-quality importance sampling, as evidenced by bidirectional techniques~\cite{Lafortune93,Veach:1994:BPT,Veach97,Keller97-IRad,Georgiev:2012:LTS}, resampling and mutation strategies~\cite{Talbot2005,Veach97,Kelemen:2002,10.1145/2766997,Hachisuka:2014:MMLT}, as well as learned distributions via neural networks~\cite{mueller2019nis,Zheng:2019,MLandIEQ} or path guiding~\cite{Jensen1995,LW1995A5TTRTVOMCRT,Vorba:2014:OnlineLearningPMMinLTS,mueller2017practical,Dahm16}.
Unfortunately, most reuse strategies come with a significant performance penalty, meaning that while they handle difficult illumination well, they are outperformed by na\"ive unidirectional path tracing under simple illumination.
This disqualifies them from the real-time setting that we strive for.
Path guiding has generally little overhead, which makes it a good choice for production rendering~\cite{Vorba2019PGP}, however adapting its underlying data structures to animated content in real-time is non-trivial ongoing work~\cite{2020_tsr_nee_pg_rtpt}.

Our importance sampling method of choice is ReSTIR~\cite{bitterli20spatiotemporal}: a recent combination of importance \emph{resampling}~\cite{Talbot2005} and classic weighted reservoir sampling~\cite{vitter85,EFRAIMIDIS2006181,chao82} that permits the reuse of a massive number of samples in constant time.
For estimating \emph{direct} lighting, ReSTIR generates many candidate samples on the light sources in the scene, resamples those candidates across space and time proportional to their predicted contribution, and then traces a shadow ray to determine the visibility of the selected sample(s).
This scheme not only has negligible overhead compared to simple path tracing, but also gracefully handles dynamic content.

In contrast, we apply ReSTIR to \emph{global} illumination by targeting it at the sum of direct illumination from light sources and indirect illumination stored in DDGI probes~\cite{Majercik2019Irradiance} of irradiance (for diffuse reflection) and angularly-filtered radiance~\cite{Majercik2021ScalingGI} (for rough glossy reflection).
This converts the global illumination problem into a purely direct illumination problem for the high variance case of low-frequency angular scattering, by treating DDGI as a light source.
In contrast to multiple importance sampling~\cite{Veach95Multiple}, this means that we draw a single sample approximately proportional to the total transported light rather than multiple samples that are heuristically combined.
Note that the result has two symbiotic levels of reuse: (i) screen-space resampling by ReSTIR for both direct and indirect light as well as (ii) world space DDGI probes which contain spatio-temporal aggregates of global illumination.

Concurrent work of Ouyang et al.~\cite{ReSTIRGI} proposes an alternative mechanism for spatio-temporal resampling of global illumination: they target ReSTIR at single-sample Monte Carlo estimates -- rather than a cache -- which can be likened to importance sampling of virtual point lights.
Compared with our approach, theirs is capable of less biased (optionally unbiased) rendering, but lacks world-space spatio-temporal reuse and early path termination, which in our case is handled by the DDGI volume.

As we show in our results, combining the underlying ideas of ReSTIR, DDGI, and path tracing into a new sampling strategy produces a less-biased and less noisy result than na\"\i{}vely compositing the results of separately-computed ReSTIR direct, DDGI diffuse/rough-glossy indirect, and path traced near-specular contributions.

\section{Algorithm}\label{sec:algorithm}

In order to render images with global illumination in
real-time, we strive to efficiently simulate radiance transport.
The rendering equation \cite{Kaj:86}
\begin{equation} \label{Eqn:REQ}
  L = L_e + T_f \; L
\end{equation}
describes the \emph{outgoing} radiance $L : \R^3 \times \mathbb{S}^2 \rightarrow \R^3$
as the sum of the radiance $L_e$ emitted by the light sources
and the transported radiance $T_f L$, where $T_f$ is a higher order operator that maps between two functions (similar to notation 
defined by Veach~\cite{VeachPhD}).
The symbol $f$ represents
the bidirectional scattering distribution function (BSDF),
describing how the surfaces in the scene transport radiance.
For our derivation, it is useful to partition radiance transport
\[
  T_f = T_{\bsdfDiff} + T_{\bsdfGlossy}
\]
into diffuse $T_{\bsdfDiff}$ and glossy $T_{\bsdfGlossy}$ transport, the latter of which including specular transport as the special case of small surface roughness.

With this operator notation at hand,
the dynamic diffuse global illumination (DDGI)
algorithm \cite{Majercik2019Irradiance} is
\begin{equation}
  L
  \approx L_e
  + T_{\bsdfDiff} \; L_e
  + L_{\text{DDGI}}
  + T_{\bsdfGlossy} \; L \;, \label{Eqn:DDGI}
\end{equation}
in which $\Lddgi \approx \sum_{i=2}^\infty (T_{\bsdfDiff})^i \; L_e$ is an approximation of the diffuse transported radiance without the directly visible light sources ($i$ starts at 2).
The other terms are estimated by sampling.
Inserting the DDGI approximation into equation~(\ref{Eqn:DDGI}) and assuming equality results almost in the rendering equation~(\ref{Eqn:REQ}), missing only higher order glossy transport.

\begin{figure}
    \small
    \centering
    \hspace*{1cm}\begin{tabularx}{\columnwidth}{XX}%
        Primary DDGI & Secondary DDGI
    \end{tabularx}
\setlength{\fboxrule}{1pt}%
\setlength{\tabcolsep}{1pt}%
\renewcommand{\arraystretch}{1}%
\small%
\centering%
\begin{tabular}{cccccccc}%

    \multicolumn{3}{c}{\setlength{\fboxsep}{1.5pt}\begin{overpic}[width=\linewidth]{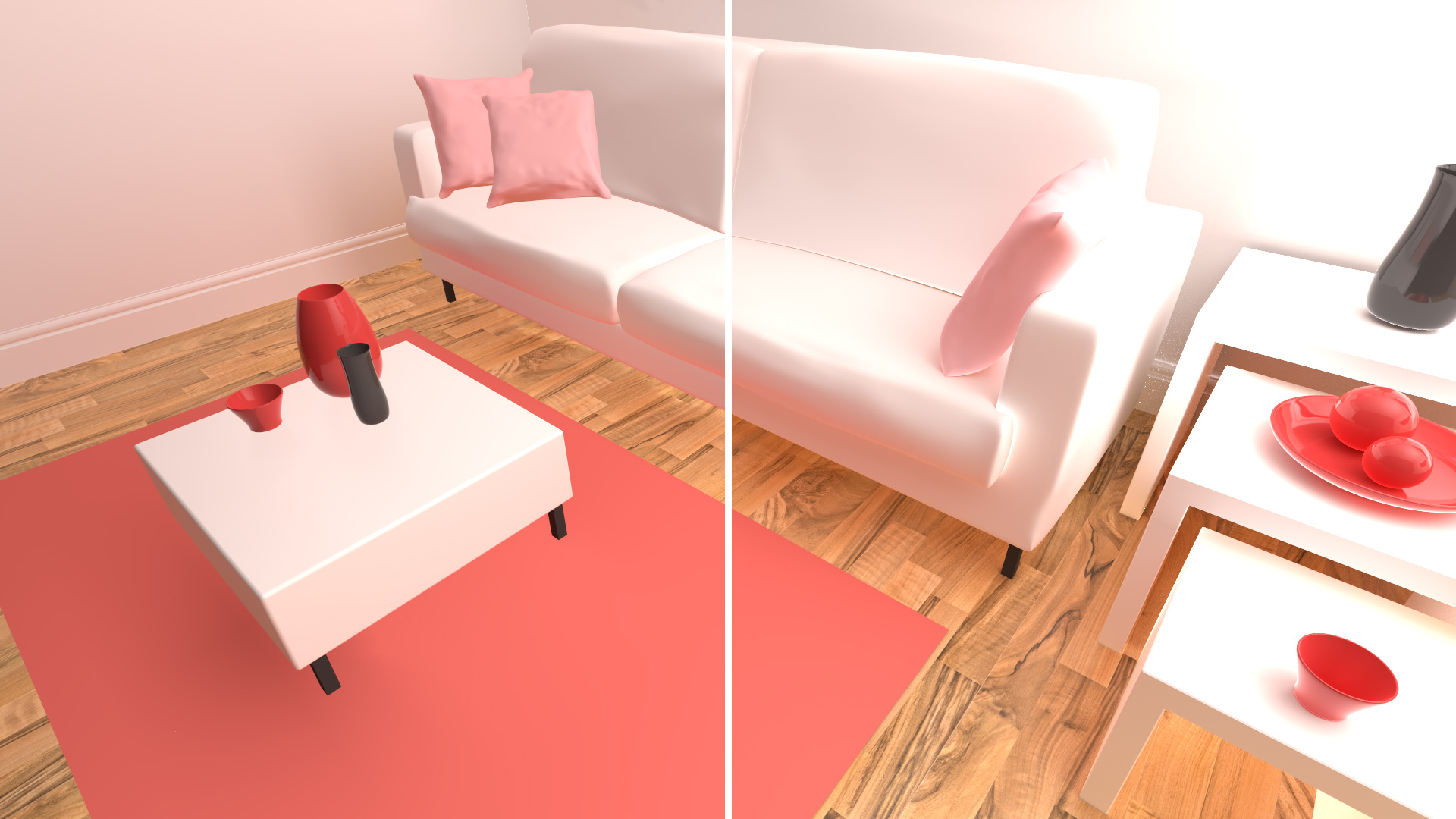}
        \put(26.5625, 28.125){\makebox(0,0){\tikz\draw[orange,ultra thick] (0,0) rectangle (0.1546875\linewidth, 0.061875\linewidth);}}
        \put(61.979166666666664, 23.177083333333332){\makebox(0,0){\tikz\draw[red,ultra thick] (0,0) rectangle (0.1546875\linewidth, 0.061875\linewidth);}}
    \end{overpic}}
    \\%
    \fcolorbox{orange}{orange}{\includegraphics[width=0.32\linewidth]{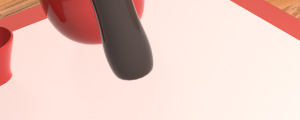}} &
    \fcolorbox{orange}{orange}{\includegraphics[width=0.32\linewidth]{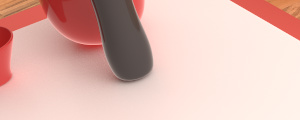}} &
    \fcolorbox{orange}{orange}{\includegraphics[width=0.32\linewidth]{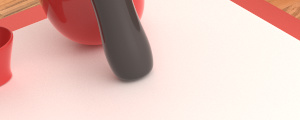}}
    \\%
    \fcolorbox{red}{red}{\includegraphics[width=0.32\linewidth]{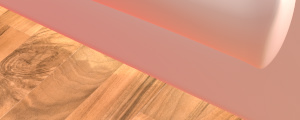}} &
    \fcolorbox{red}{red}{\includegraphics[width=0.32\linewidth]{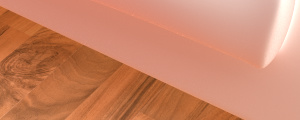}} &
    \fcolorbox{red}{red}{\includegraphics[width=0.32\linewidth]{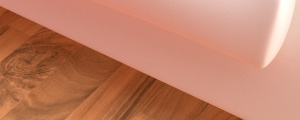}}
    \\%
            Primary DDGI
                    &
                    Secondary DDGI
                    & Reference
                \\%
\end{tabular}
    \caption{Using the DDGI approximation $\Lddgi$ at the primary vertex (left) shows light leaking artifacts beneath the couch and does not correctly reproduce indirect contact shadows on the table, vases, and plate. Using our augmented $\LddgiP$ at the \emph{secondary} vertex (middle) exhibits no leaks, recovers some of the indirect contact shadows and is closer to the path traced reference (right).
    }\label{fig:occlusionComparison}
\end{figure}

Majercik et al.~\cite{Majercik2021ScalingGI} extended DDGI to include this missing contribution for the special case of specular reflection, which we will build on in \autoref{Sec:Glossy}.
The ``Primary DDGI'' column in Fig.~\ref{fig:occlusionComparison} and \ref{fig:main-result1spp}
shows results of this extended DDGI algorithm. As compared to path tracing
\cite{Kaj:86}, the images are smooth and noise only results
from sampling the direct diffuse illumination $T_{\bsdfDiff} \; L_e$ and the recursion for glossy transport $T_{\bsdfGlossy} \; L$.
Yet, the irradiance probe approximation of $L_{\text{DDGI}}$
shows visible artifacts that we resolve by postponing its evaluation to secondary path vertices.
We develop our new algorithm in the following sections.

\begin{figure}
    \small
    \includegraphics[width=\linewidth]{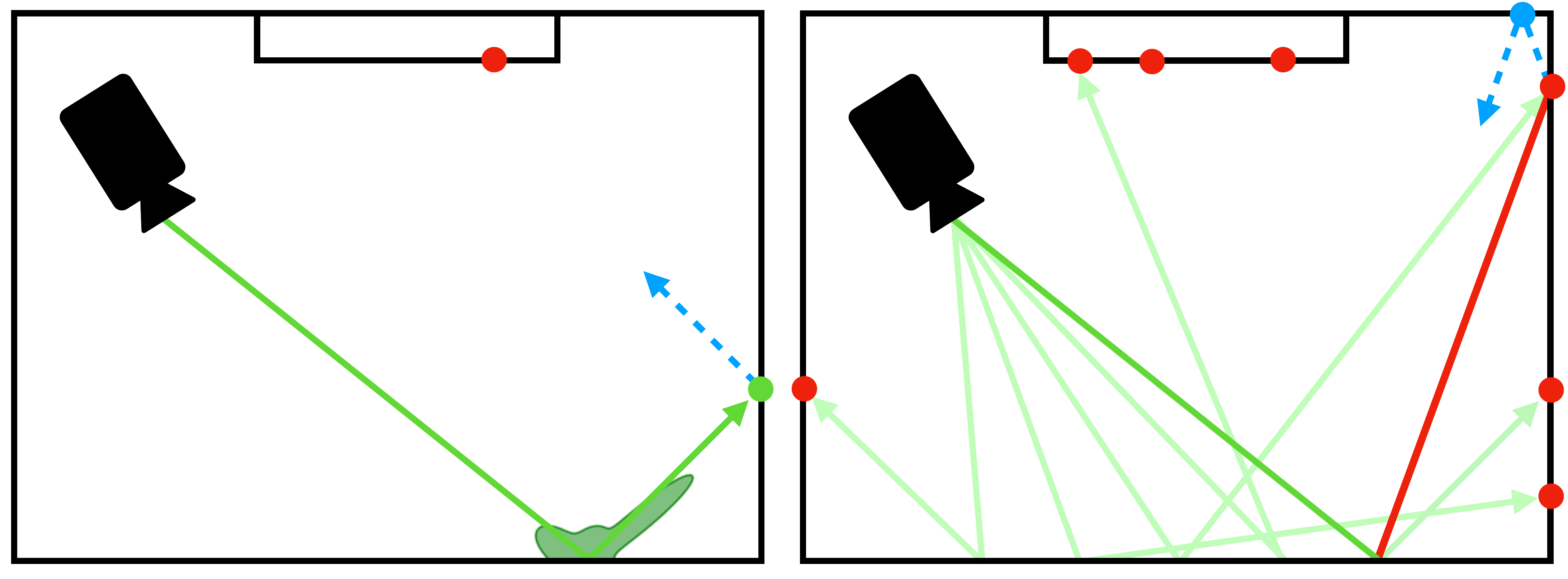}%
    \setlength{\unitlength}{0.01\linewidth}%
    \begin{picture}(0,0)
        \put(-62.5, -2) {$\pos$}
        \put(-55.5, 11.5) {$\posy$}
        \put(-13, -2) {$\pos$}
        \put(1, 30.5) {$\posz$}
    \end{picture}
    \caption{We compute direct and indirect illumination in a unified way by targeting the ReSTIR algorithm at the sum of emitted light $L_e$ and the DDGI approximation $\Lddgi$.
    Left: two points are sampled. One (red) is sampled on the emissive surfaces and another (green, $\posy$) by BSDF sampling at the primary vertex $\pos$.
    Right: with all samples across pixels and past frames converted to the area measure, a single sample $\posz$ is selected with probability proportional to ${L_e+\Lddgi}$ 
    and a shadow ray (red) is traced.
    The radiance leaving $\pos$ towards the camera is estimated with low variance by combining the reflected light from $\posy$ and $\posz$ through the method of Kollig and Keller~\cite{Kollig2006Singularity}.
    (Near-)specular transport at $\posy$ and $\posz$ is estimated through recursion (blue) and the remaining transport by $\Lddgi$.
}\label{fig:diagrams}
\end{figure}

\subsection{Sampling all Surfaces with ReSTIR} \label{Sec:Sampling}

As the DDGI approximation can be queried at any point
of the scene surface, we consider the whole scene surface as a
light source rather than only the actual light sources.
Our new sampling algorithm performs the following steps:
\begin{enumerate}
    \item Generate candidate samples on {\em all} surfaces of the scene.
    To this end, we uniformly sample a small number of positions on emissive surfaces and generate additional samples by tracing secondary rays in directions sampled from the BSDF.
    \item Select one candidate by first sampling from a weighted reservoir proportional to the sample contribution, then
    resample spatially among neighboring pixels, and resample temporally among previous frames, proportional to the sum ${L_e + \Lddgi}$.
    \item Trace one shadow ray to the selected sample point.
    \item If visible, shade from that sample point, using the emissive contribution $L_e$ and adding the contribution
    from the DDGI volume.
\end{enumerate}
As with original ReSTIR, our unified sampling scheme traces only a single ray to the selected sample (see Figure~\ref{fig:diagrams}).
However, as the original DDGI volume only contains the diffuse
\emph{indirectly reflected} light without direct reflections as shown in equation~(\ref{Eqn:DDGI}),
we need to add these, as well as glossy contributions, which we describe next.

\subsection{Augmenting the DDGI Approximation} \label{Sec:Augmentation}

As mentioned in Sec.~\ref{Sec:Sampling},
integrating $L_{\text{DDGI}}$ from secondary vertices over the hemisphere at the primary vertex
would result in too dark images, because
the DDGI approximation lacks the diffuse direct illumination
and the radiance from non-diffuse transport.
In fact, adding these missing terms to the DDGI approximation recovers an approximation of the transport operator
\begin{equation} \label{Eqn:DGI}
  L_{\text{DDGI}} + T_{\bsdfDiff} \; L_e + T_{\bsdfGlossy} \; L
  \approx T_{\bsdfDiff} \; L + T_{\bsdfGlossy} \; L
  = T_f \; L .
\end{equation}

Adding the missing terms
by recursive path tracing would increase the noise and involve
further shading and tracing cost.

Instead, we augment the DDGI approximation in two steps, which are most easily explained by examining the underlying irradiance approximation that DDGI makes.
Given a 3D position and 2D normal vector, the DDGI probe volume stores irradiance from indirect diffuse reflections, which is approximated by repeatedly evaluating the following double-bounce transport~\cite{Majercik2019Irradiance}
\begin{align}
    \Eddgi(\pos, \normal) &\approx \int (\Lddgi + T_{\bsdfDiff} \; L_e)(h(\pos, \diri), -\diri) \langle \normal, \diri \rangle \,\Diff{\diri} \,, \nonumber \\
    \Lddgi(\pos, \diri) &:= \frac{\rho(\pos)}{\pi} \; \Eddgi(\pos, \normal) \,,
\end{align}
where $h$ is the ray tracing operation and $\rho$ is the diffuse albedo.
Repeated Monte Carlo estimation of this equation at the centers $\pos$ of all DDGI probes converges to the aforementioned approximation $$\Lddgi \approx \sum_{i=2}^\infty (T_{\bsdfDiff})^i \; L_e \,.$$
To include direct reflections and non-diffuse transport, we replace the double-bounce transport with simpler single-bounce recursion
\begin{align}
    \EddgiP(\pos, \normal) \approx \int (\LddgiP + L_e)(h(\pos, \diri), -\diri) \langle \normal, \diri \rangle \,\Diff{\diri} \,,
    \label{eq:probeUpdate}
\end{align}
where $L_e$ no longer undergoes a diffuse interaction and $\LddgiP$ will be adapted to include glossy transport in the following section.
The approximation thus converges to the full transport operator
$$\LddgiP \approx T_{\bsdf} \; L \,.$$

Including direct illumination (i.e.\ untransported $L_e$) in the probe update equation~(\ref{eq:probeUpdate}) produces results that are too bright whenever probe centers are closer to light sources than the surfaces that they shade.
This causes surfaces to receive more (secondary) direct illumination than they should -- a common problem in probe-based caches.
To alleviate this, we approximate the true distance to the surface by querying the average visibility already computed by the probes in the \emph{backwards} direction~\cite{Majercik2019Irradiance} (dashed red lines), clipping that value to the probe boundary (pink), and adding it to the length of the corresponding probe update rays (solid red arrows). Because the update rays are traced per probe, this costs just a texture read without any probe weight computation.
\begin{wrapfigure}{r}{0.4\linewidth}
    \small
    \vspace{-1mm}\hspace*{-3mm}\includegraphics[width=1.1\linewidth]{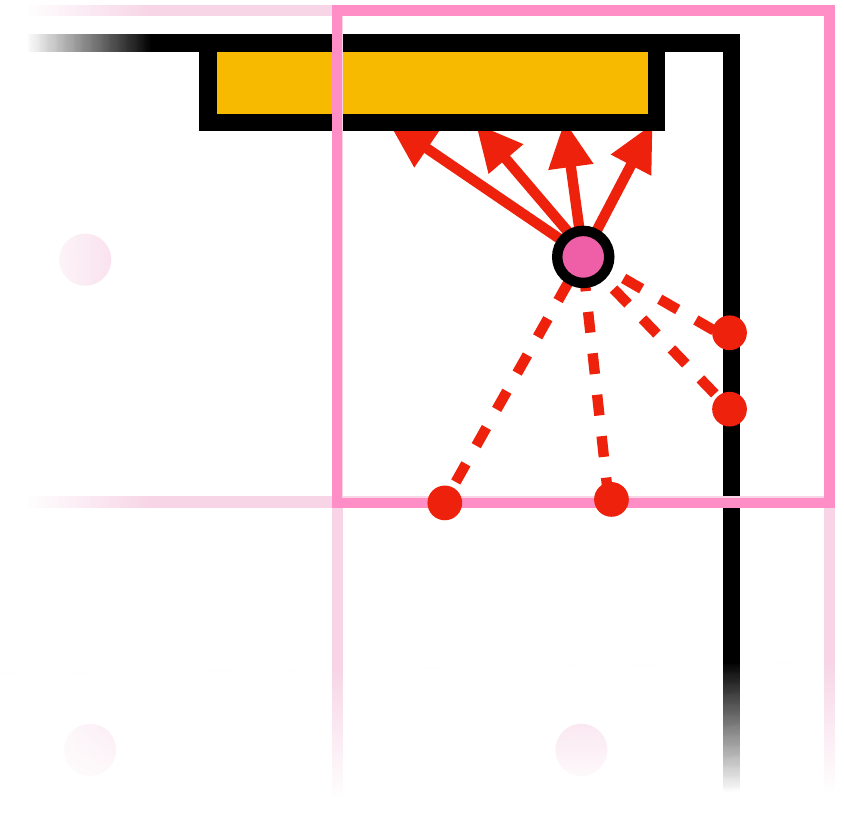}%
    \setlength{\unitlength}{0.011\linewidth}%
    \begin{picture}(0,0)
    \end{picture}\\[-2mm]
\end{wrapfigure}
We then use the inverse square of the \emph{total} distance, i.e.,\ solid + dashed, to attenuate $L_e$.
In contrast to direct illumination computed at probe centers, this heuristic slightly underestimates direct illumination.
We therefore expose a user-tunable slider to scale the additional attenuation per scene to correct for brightness differences that would otherwise be a significant source of bias.

\begin{figure}
    \centering
    \vspace{12mm}
\setlength{\fboxrule}{1pt}%
\setlength{\tabcolsep}{1pt}%
\renewcommand{\arraystretch}{1}%
\small%
\centering%
\vspace{-10mm}
\begin{tabular}{cccccccc}%

    \multicolumn{3}{c}{\setlength{\fboxsep}{1.5pt}\begin{overpic}[width=\linewidth]{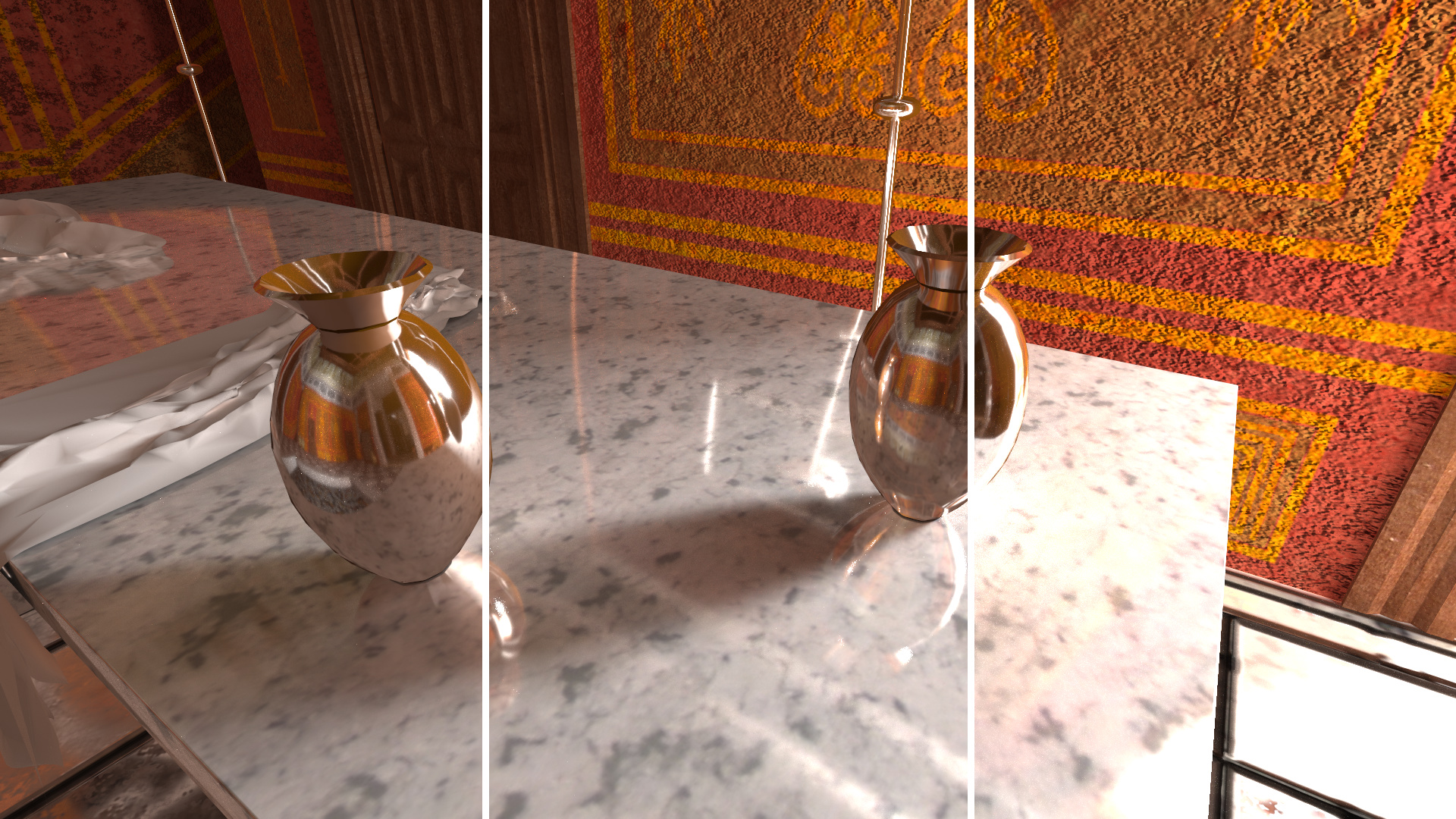}
        \put(25.3125, 33.854166666666664){\makebox(0,0){\tikz\draw[red,ultra thick] (0,0) rectangle (0.10312500000000001\linewidth, 0.041249999999999995\linewidth);}}
        \put(31.354166666666668, 16.145833333333332){\makebox(0,0){\tikz\draw[orange,ultra thick] (0,0) rectangle (0.10312500000000001\linewidth, 0.041249999999999995\linewidth);}}
    \end{overpic}}
    \\%
    \fcolorbox{red}{red}{\includegraphics[width=0.32\linewidth]{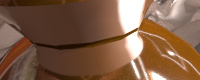}} &
    \fcolorbox{red}{red}{\includegraphics[width=0.32\linewidth]{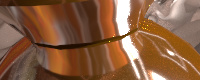}} &
    \fcolorbox{red}{red}{\includegraphics[width=0.32\linewidth]{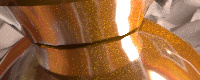}}
    \\%
    \fcolorbox{orange}{orange}{\includegraphics[width=0.32\linewidth]{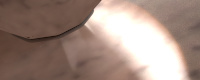}} &
    \fcolorbox{orange}{orange}{\includegraphics[width=0.32\linewidth]{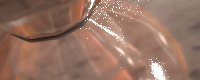}} &
    \fcolorbox{orange}{orange}{\includegraphics[width=0.32\linewidth]{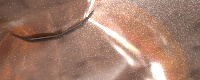}}
    \\%
            Secondary DDGI
                    &
                    +Specular Recursion
                    & Reference
                \\%
\end{tabular}
    \caption{For shading, we recursively trace (near-)specular interactions until their footprint grows sufficiently large to blur away small-scale artifacts of the DDGI approximation (middle).
    This produces less biased results than always querying the DDGI approximation at the secondary vertex (left).}
    \label{fig:glossyDDGI}
    \vspace{-4mm}
\end{figure}

\subsection{Including Glossy Illumination} \label{Sec:Glossy}

Recall that the reflected light from a diffuse material with albedo $\rho$ can be cheaply approximated by looking up $\frac{\rho(\pos)}{\pi} \;\EddgiP(\pos, \normal)$.

Majercik et al.~\cite{Majercik2021ScalingGI} make the observation that the irradiance approximation can \emph{additionally} be used
to very coarsely approximate specular transport by substituting the normal $\normal$ with the direction of specular reflection $\dirr$.
The irradiance $\Eddgi(\pos, \dirr)$ can then be \emph{re-interpreted} as a prefiltered (by the cosine term) coarse approximation of incident radiance up to a normalization factor of $2\pi$ which must be added.
In the illustration, the green lobes visualize cosine-weighted prefiltering of incident radiance and the red arrows indicate the query direction of $\Eddgi$.

\begin{figure}[h]
    \small
    \centering
    \vspace{-1mm}\hspace*{0mm}\includegraphics[width=0.8\linewidth]{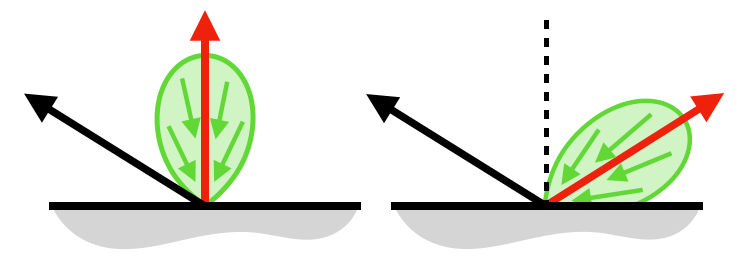}%
    \setlength{\unitlength}{0.008\linewidth}%
    \begin{picture}(0,0)
        \put(-73, 4) {$\pos$}
        \put(-70, 30) {$\normal$}
        \put(-28, 4) {$\pos$}
        \put(-25, 30) {$\normal$}
        \put(-68, 10) {$\diri$}
        \put(-8, 10) {$\diri$}
        \put(-2, 20) {$\dirr$}
        \put(-90, 20) {$\diro$}
        \put(-44, 20) {$\diro$}
        \put(-82, -5) { Diffuse }
        \put(-37, -5) { Specular }
    \end{picture}\\[2mm]
\end{figure}

While Majercik et al.\ use this strategy for just specular transport, we generalize it to arbitrary glossy BSDFs -- in our case microfacet models -- by Monte Carlo estimation.
More specifically, we approximate reflected radiance by our augmented DDGI model
\begin{align}
    \LddgiP(\pos, \diro) \label{Eqn:lddgiplus}
    &= \underbrace{\frac{\rho(\pos)}{\pi} \; \EddgiP(\pos, \normal)}_{\text{Diffuse}}\\
    &+ \underbrace{\frac{1}{2\pi} \int \EddgiP(\pos, \diri) \bsdfGlossy(\pos, \diro, \diri) \langle \normal, \diri \rangle \,\Diff{\diri} }_{\text{Glossy}} \nonumber
\end{align}
as the decomposition into diffuse and glossy transport, where we estimate the glossy integral with a single Monte Carlo sample drawn proportionally to the glossy BSDF component $\bsdfGlossy$.

\paragraph*{Specular recursion for shading.}
We use our radiance approximaton $\LddgiP$ in three parts of the algorithm: (i) in the update rule of $\EddgiP$~(\ref{eq:probeUpdate}), (ii) to resample proportional to ${(L_e + \LddgiP)}$, and (iii) for shading.
The latter use case -- shading -- requires special care to avoid exposing too much of the inherent bias of $\LddgiP$ through (near-)specular interactions.

We thus trace shading paths recursively until their scattering footprint is sufficiently spread out to blur DDGI's visual artifacts.
To this end, we adopt the path termination strategy of Müller et al.~\cite{mueller2021realtime,mueller2020neural}, which is based on the isotropic path footprint approximation of Bekaert et al.~\cite{bekaert2003}.

More specifically, we approximate the footprint of a path with vertices $\pos_1\cdots\pos_n$ as
\begin{align}
    a(\pos_1\cdots\pos_n) &= {\left( \sum_{i=1}^{n-1} \sqrt{\frac{\|\pos_{i} - \pos_{i+1}\|^2}{p_i(\dir_{\mathrm{i},i} \,|\,\pos_{i}, \dir_{\mathrm{o},i}) \, \langle \normal_{i+1}, -\dir_{\mathrm{i},i} \rangle}} \right)}^2 \,,
\end{align}
where $p_i$ is the BSDF sampling density at the $i$-th vertex.
At the primary vertex, $p_1$ is proportional to the entire BSDF $\bsdf$, whereas ${p_i; \, i>1}$ is proportional to only the glossy portion of the BSDF $\bsdfGlossy$, because we simply terminate the recursion early and query $\frac{\rho(\pos_n)}{\pi}\;\EddgiP(\pos_n, \normal_n)$ when the diffuse portion is sampled.
When the glossy portion is sampled, we terminate the recursion by estimating the glossy integral of Eq.~(\ref{Eqn:lddgiplus}) as soon as $a(\pos_1\cdots\pos_n) > a_0 \cdot c$, where
\begin{align}
    \AreaHeuristic_0 := \frac{\|\pos_0 - \pos_1\|^2}{4\pi \; \langle \normal_1, -\dir_{\mathrm{i},i} \rangle}
\end{align}
is an approximation (up to constant factors) of the pixel footprint projected into the scene, $\pos_0$ is the camera position, and $c$ is a user-chosen threshold.
See Bekaert et al.~\cite{bekaert2003} and Müller et al.~\cite{mueller2021realtime,mueller2020neural} for details.

We empirically set the threshold ${c = 0.2}$ to obtain a satisfactory noise versus bias trade-off.
Fig.~\ref{fig:glossyDDGI} confirms that this helps reduce artifacts on highly glossy surfaces without affecting rough glossy and diffuse surfaces.

\subsection{Correcting Self-shadow Bias} \label{Sec:SelfShadowBias}

\begin{figure}
\centering
\vspace{-1mm}
\setlength{\fboxrule}{10pt}%
\setlength{\insetvsep}{20pt}%
\setlength{\tabcolsep}{-0.1pt}%
\renewcommand{\arraystretch}{1}%
\footnotesize%
\hspace*{0.5mm}\begin{tabular}{lccccccc}
    &
    &Default SS Bias
    &View Path Bias
    \\%
    \setInset{A}{red}{375}{950}{244}{120}%
    \setInset{B}{orange}{700}{882}{244}{120}%
    &\addBeautyCrop{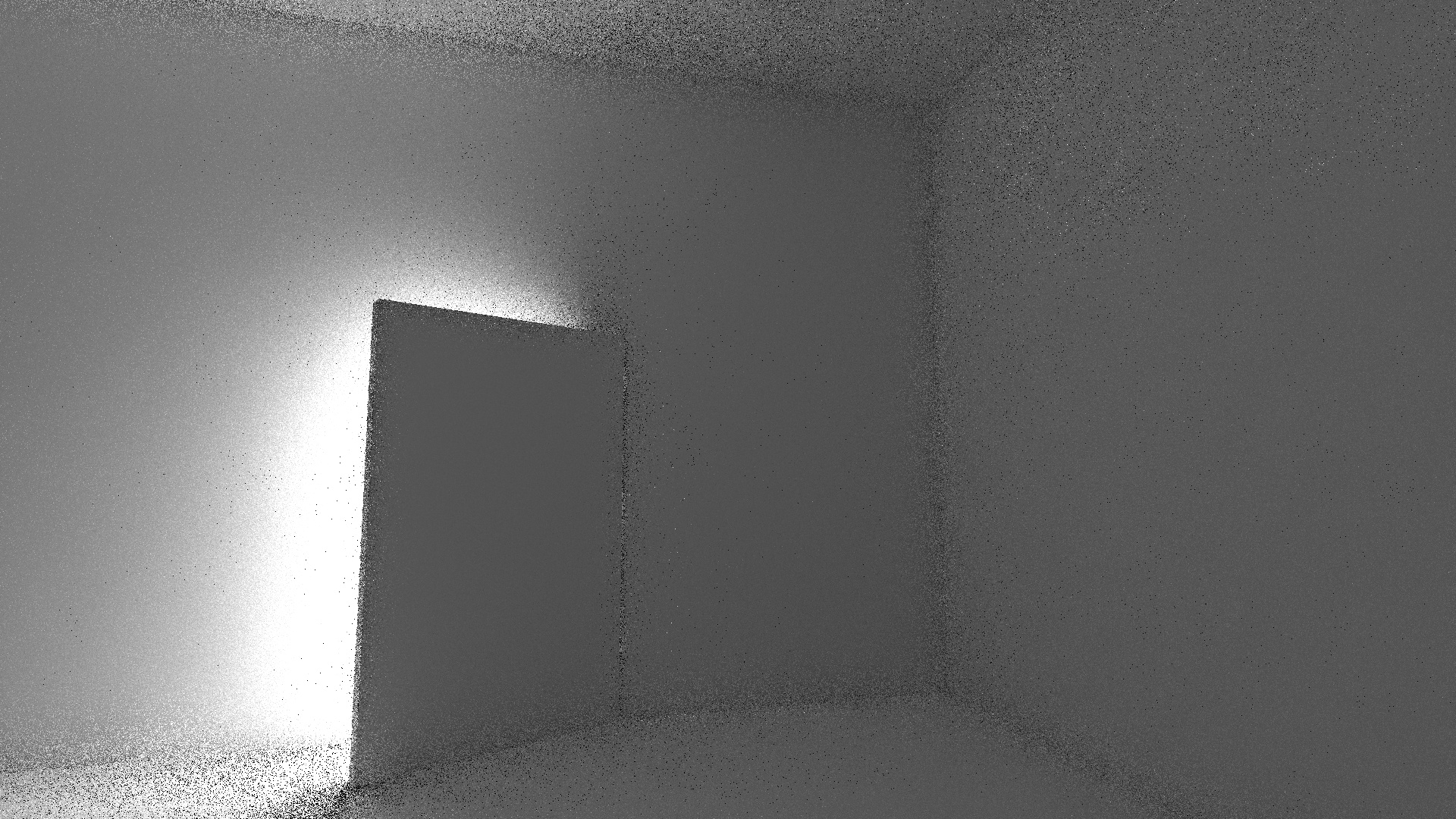}{0.23}{1920}{1080}{0}{0}{1920}{1080}%
    &\addInsets{generated_figures/fig_ss_bias_1/room_door/Default-Bias.jpg}%
    &\addInsets{generated_figures/fig_ss_bias_1/room_door/View-Path-Bias.jpg}%
    & 
    \\%
\end{tabular}
\includegraphics[width=1.015\linewidth]{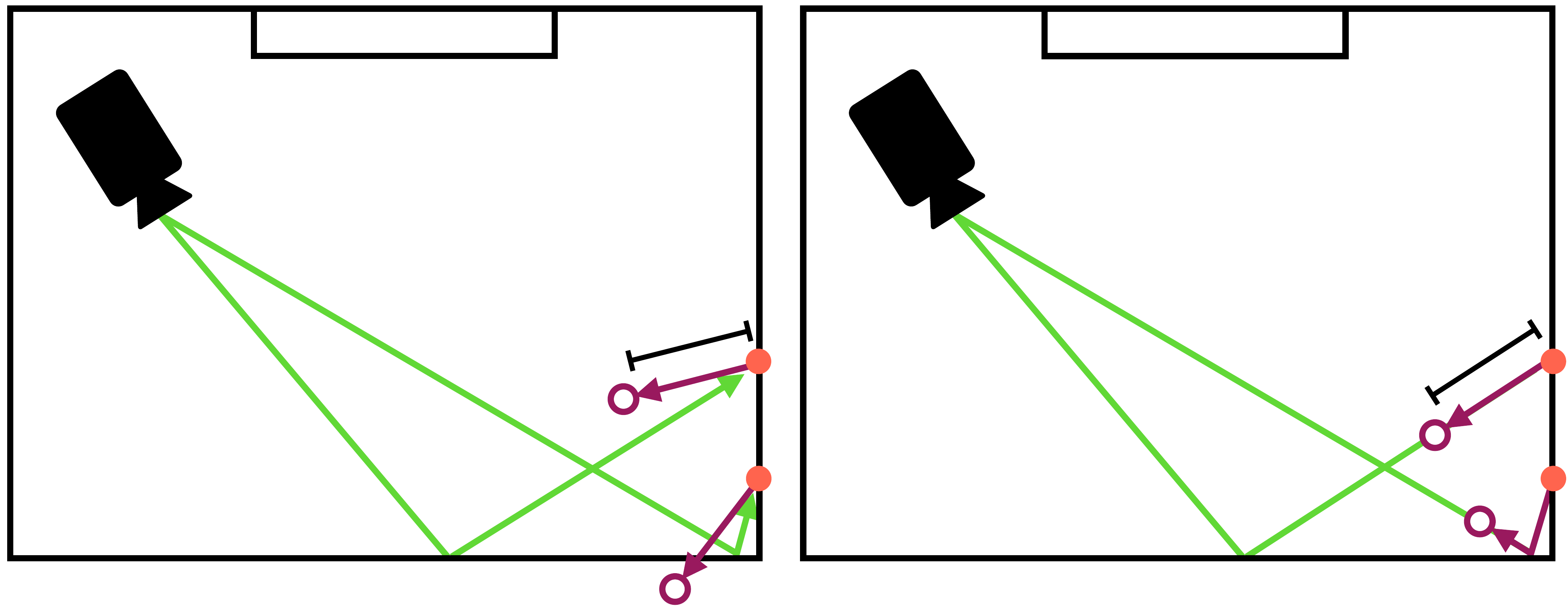}
\setlength{\unitlength}{0.01\linewidth}%
    \begin{picture}(0,0)
        \put(43, 23) {$d$}
        \put(93.5, 21.5) {$d$}
    \end{picture}
    \begin{subfigure}{0.49\linewidth}
        \caption{Default DDGI self-shadow bias }
    \end{subfigure}
    \begin{subfigure}{0.49\linewidth}
        \caption{New view path bias}
    \end{subfigure}
    \vspace{-3mm}
    \caption{DDGI combats self-shadowing by pushing the evaluation point away from surfaces (violet arrow) by a fixed distance $d$ that is proportional to the probe size. Left: DDGI uses a linear combination of the view direction and the surface normal.
    This may push the evaluation point through geometry, leading to light leaking.
    Right: we instead move the evaluation point a fixed distance back along the path, which is guaranteed to not penetrate surfaces.}\label{fig:ssbiasdiagram}
    \vspace{-1mm}
\end{figure}

The original DDGI algorithm applies an offset to the positions at which $\Lddgi$ is queried to avoid computing visibility weights exactly at a surface where their variance is high. This offset is called the ``self-shadow'' bias, and is computed as a weighted combination of the view and normal vector at the queried point. 

While the self-shadow bias was sufficient for the original primary-vertex DDGI algorithm, using it unmodified in our indirect sampling strategy showed light leaking artifacts (Figure~\ref{fig:ssbiasdiagram}). These artifacts arise from very short distances between the view point and the query point, as might arise from sharp corners. To address these artifacts, we bias the query point along the \emph{view path} as opposed to the surface normal and view direction. Results using our view path bias against the original self-shadow bias are shown in Figure~\ref{fig:ssbiasdiagram}.

\subsection{Taming the Weak Singularity in the Geometry Term}%
\label{Sec:Singularity}

\begin{figure}
    \centering
\setlength{\fboxrule}{1pt}%
\setlength{\tabcolsep}{1pt}%
\renewcommand{\arraystretch}{1}%
\small%
\centering%
\begin{tabular}{cccccccc}%

    \multicolumn{3}{c}{\setlength{\fboxsep}{1.5pt}\begin{overpic}[width=\linewidth]{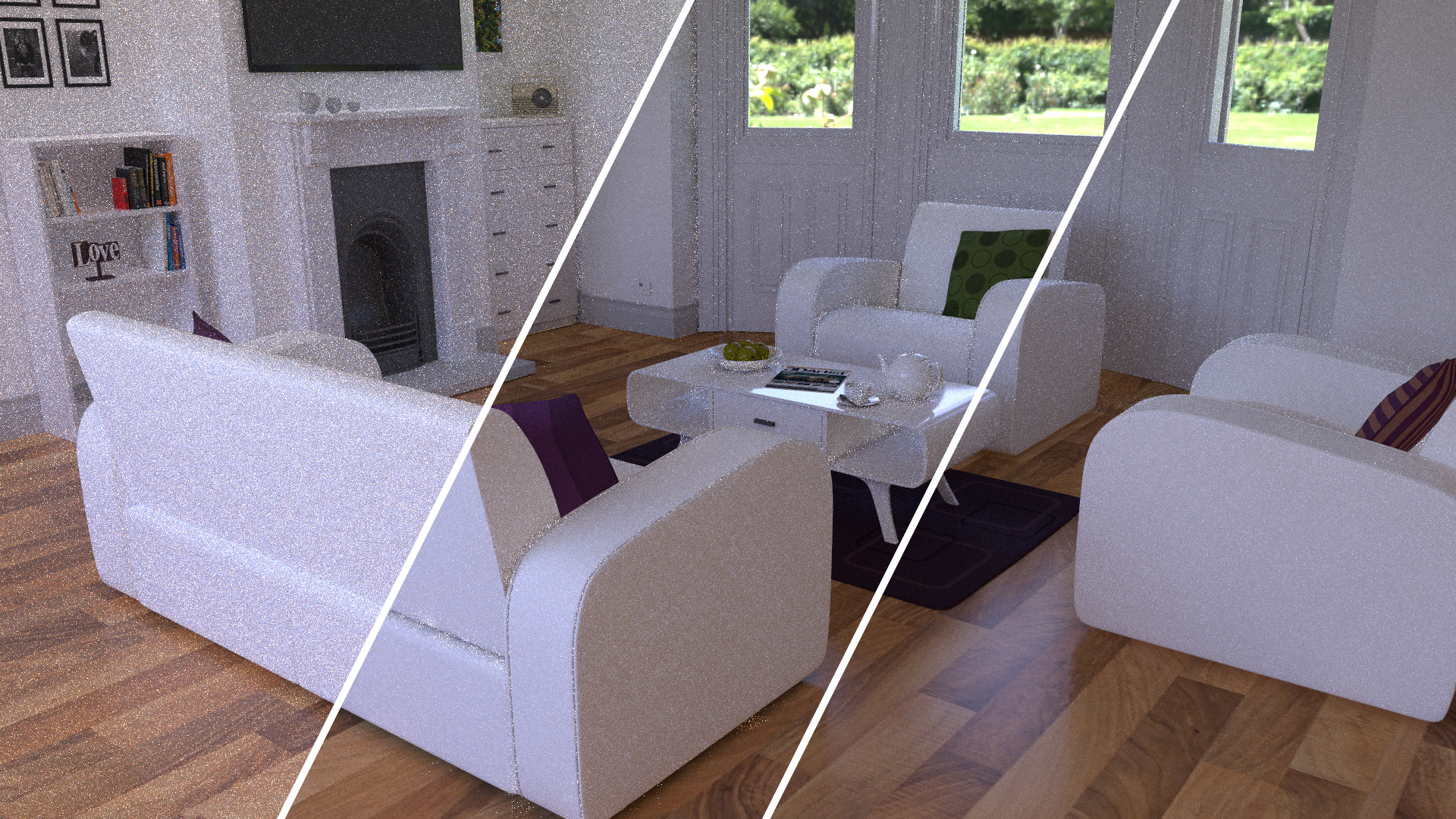}
        \put(28.072916666666664, 14.0625){\makebox(0,0){\tikz\draw[orange,ultra thick] (0,0) rectangle (0.10312500000000001\linewidth, 0.041249999999999995\linewidth);}}
        \put(58.90625000000001, 34.895833333333336){\makebox(0,0){\tikz\draw[red,ultra thick] (0,0) rectangle (0.10312500000000001\linewidth, 0.041249999999999995\linewidth);}}
    \end{overpic}}
    \\%
    \fcolorbox{orange}{orange}{\includegraphics[width=0.32\linewidth]{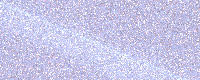}} &
    \fcolorbox{orange}{orange}{\includegraphics[width=0.32\linewidth]{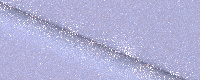}} &
    \fcolorbox{orange}{orange}{\includegraphics[width=0.32\linewidth]{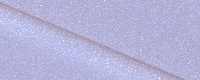}}
    \\%
    \fcolorbox{red}{red}{\includegraphics[width=0.32\linewidth]{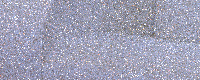}} &
    \fcolorbox{red}{red}{\includegraphics[width=0.32\linewidth]{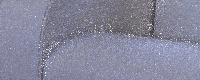}} &
    \fcolorbox{red}{red}{\includegraphics[width=0.32\linewidth]{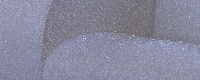}}
    \\%
            PT + ReSTIR
                    &
                    DDGI Resampling
                    &
                    DDGI Resampling\\
                    &
                    without bounding
                    &
                    with bounding (ours)
                \\%
\end{tabular}
    \caption{Even though spatio-temporal resampling of DDGI (middle) improves the noise of path tracing (left), outliers (``fireflies'') cause distracting artifacts in geometric concavities.
    These fireflies are due to the weak singularity in the geometry term of the rendering equation. Using the method of Kollig and Keller~\cite{Kollig2006Singularity}, we bound the geometry term during resampling and then trace an auxiliary ray to estimate the residual transport (right). This eliminates fireflies and preserves a significant noise reduction. The images were rendered at 128 samples per pixel.
}
    \label{fig:geometryBounding}
\end{figure}

In order to perform spatio-temporal resampling among different shading locations, the geometry term is introduced as part of the change of the integration measure.
Unfortunately, the weak singularity in the geometry term may lead to unbounded variance in the resampling procedure in geometric creases; see Fig.~\ref{fig:geometryBounding} (middle).

In typical path tracing applications, uniform multiple importance sampling~\cite{Veach95Multiple} of light sources would bound the variance when light sources (almost) touch the surfaces that they shade.
But this is an impractical solution in our case, where the entire scene is considered a light source.
It may be feasible to \emph{non-uniformly} sample all relevant scene surfaces according to a data structure that is constructed online, but we instead take a simpler approach that does not require multiple importance sampling: we partition the integrand to bound the variance~\cite{Kollig2006Singularity}.

To avoid the weak singularity caused by samples very close to the primary vertex, we clamp the computed geometry term to a maximum value of $G_\text{max} = 1$ and use the notation
$\lceil T_{\bsdf} \rceil$ for the
transport with this bounded geometry term.
The residual transport is denoted by $\lfloor T_{\bsdf} \rfloor$,
such that
\[
  T_{\bsdf} = \lceil T_{\bsdf} \rceil + \lfloor T_{\bsdf} \rfloor .
\]
$\lfloor T_{\bsdf} \rfloor$ is evaluated by
an auxiliary BSDF sample in the sampling step described in Sec.~\ref{Sec:Sampling}.
The improved variance is shown in Fig.~\ref{fig:geometryBounding}.

\begin{figure}
    \centering
\setlength{\fboxrule}{1pt}%
\setlength{\tabcolsep}{1pt}%
\renewcommand{\arraystretch}{1}%
\small%
\centering%
\begin{tabular}{cccccccc}%

        \multicolumn{3}{c}{\setlength{\fboxsep}{1.5pt}\begin{overpic}[width=\linewidth]{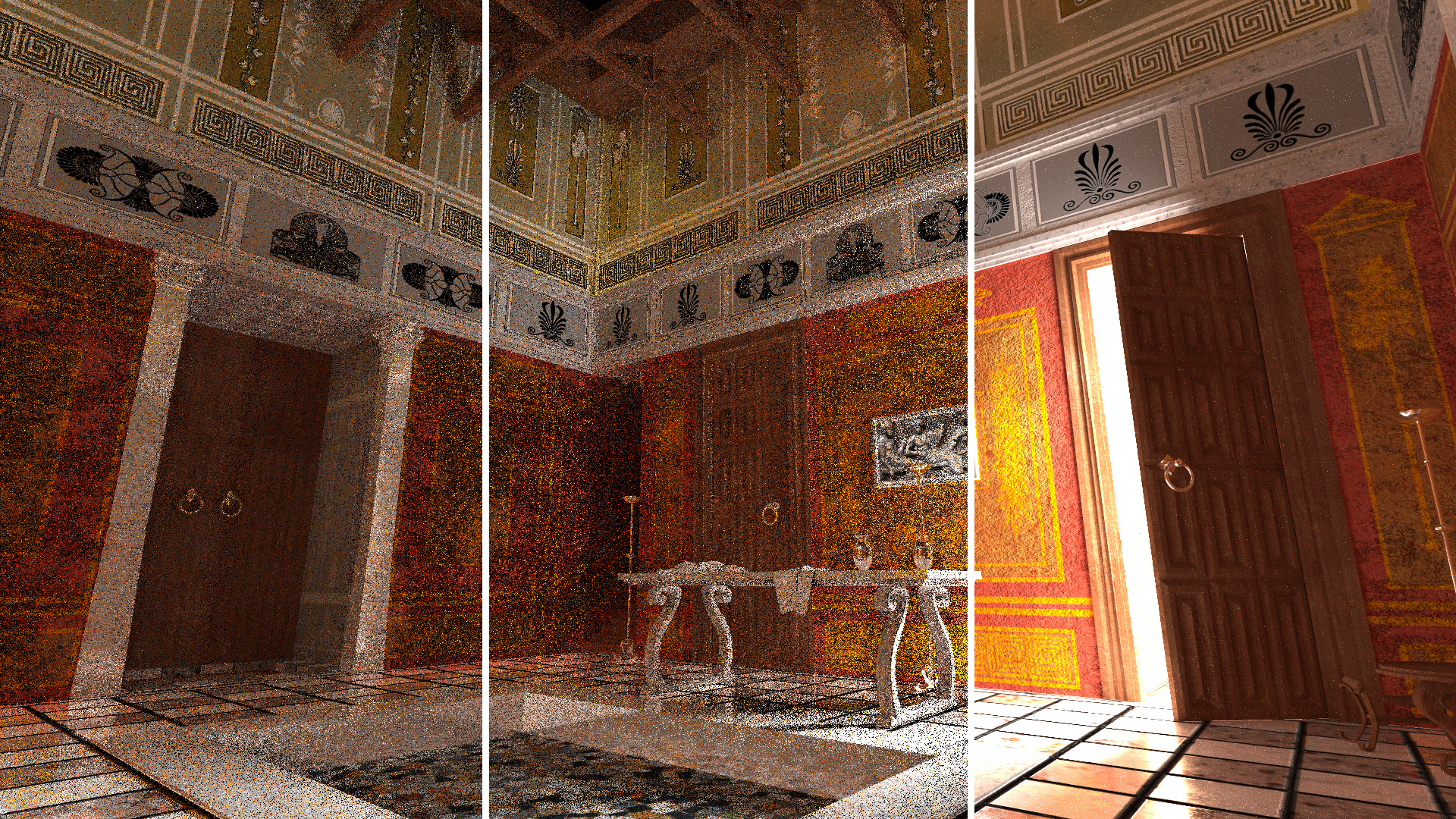}
        \put(38.54166666666667, 48.17708333333333){\makebox(0,0){\tikz\draw[red,ultra thick] (0,0) rectangle (0.10312500000000001\linewidth, 0.041249999999999995\linewidth);}}
        \put(35.104166666666664, 35.416666666666664){\makebox(0,0){\tikz\draw[orange,ultra thick] (0,0) rectangle (0.10312500000000001\linewidth, 0.041249999999999995\linewidth);}}
    \end{overpic}}
    \\%
    \fcolorbox{red}{red}{\includegraphics[width=0.32\linewidth]{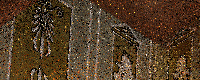}} &
    \fcolorbox{red}{red}{\includegraphics[width=0.32\linewidth]{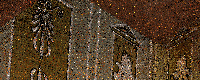}} &
    \fcolorbox{red}{red}{\includegraphics[width=0.32\linewidth]{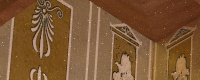}}
    \\%
    \fcolorbox{orange}{orange}{\includegraphics[width=0.32\linewidth]{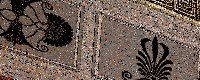}} &
    \fcolorbox{orange}{orange}{\includegraphics[width=0.32\linewidth]{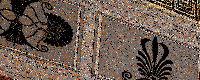}} &
    \fcolorbox{orange}{orange}{\includegraphics[width=0.32\linewidth]{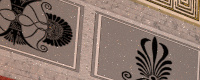}}
    \\%
            DDGI Resampling
                    &
                    + Query Caching
                    & Reference
                \\%

            MAPE: 0.63 &
            MAPE: 0.64 &
        \\%

            50.3 ms &
            18.4 ms &
        \\%
\end{tabular}
    \caption{Caching the DDGI query for resampling is more efficient: while the increase in noise level and MAPE is negligible in query caching, the reduction in frame rendering time is substantial.}
    \label{fig:caching}
    \vspace{-4mm}
\end{figure}

In summary, our new algorithm to approximate the rendering equation
amounts to summing separate estimators of each term of
\begin{equation}
  L
  \approx L_e 
   + \underbrace{\lceil T_{\bsdf} \rceil \; (L_e + \LddgiP)}_{\text{bounded geometry term}}
  + \underbrace{\lfloor T_{\bsdf} \rfloor \; (L_e + \LddgiP)}_{\text{residual transport}} , \label{Eqn:Algorithm}
\end{equation}
where
\begin{itemize}
    \item $L_e$ is evaluated at the primary vertex,
    \item $\lfloor T_{\bsdf} \rfloor$ by importance sampling the BSDF, and
    \item $\lceil T_{\bsdf} \rceil$ by spatio-temporal reservoir resampling.
\end{itemize}

\subsection{Improved Performance by Caching DDGI Queries}\label{sec:caching}

During the spatio-temporal resampling step, our algorithm requires numerous DDGI queries:
one for each candidate's contribution to the shaded pixel, which, in our implementation, amounts to 1 temporal and 3 spatial candidates per pixel.
\begin{wrapfigure}{r}{0.5\linewidth}
    \small
    \vspace{-1mm}\hspace*{-5mm}\includegraphics[width=1.1\linewidth]{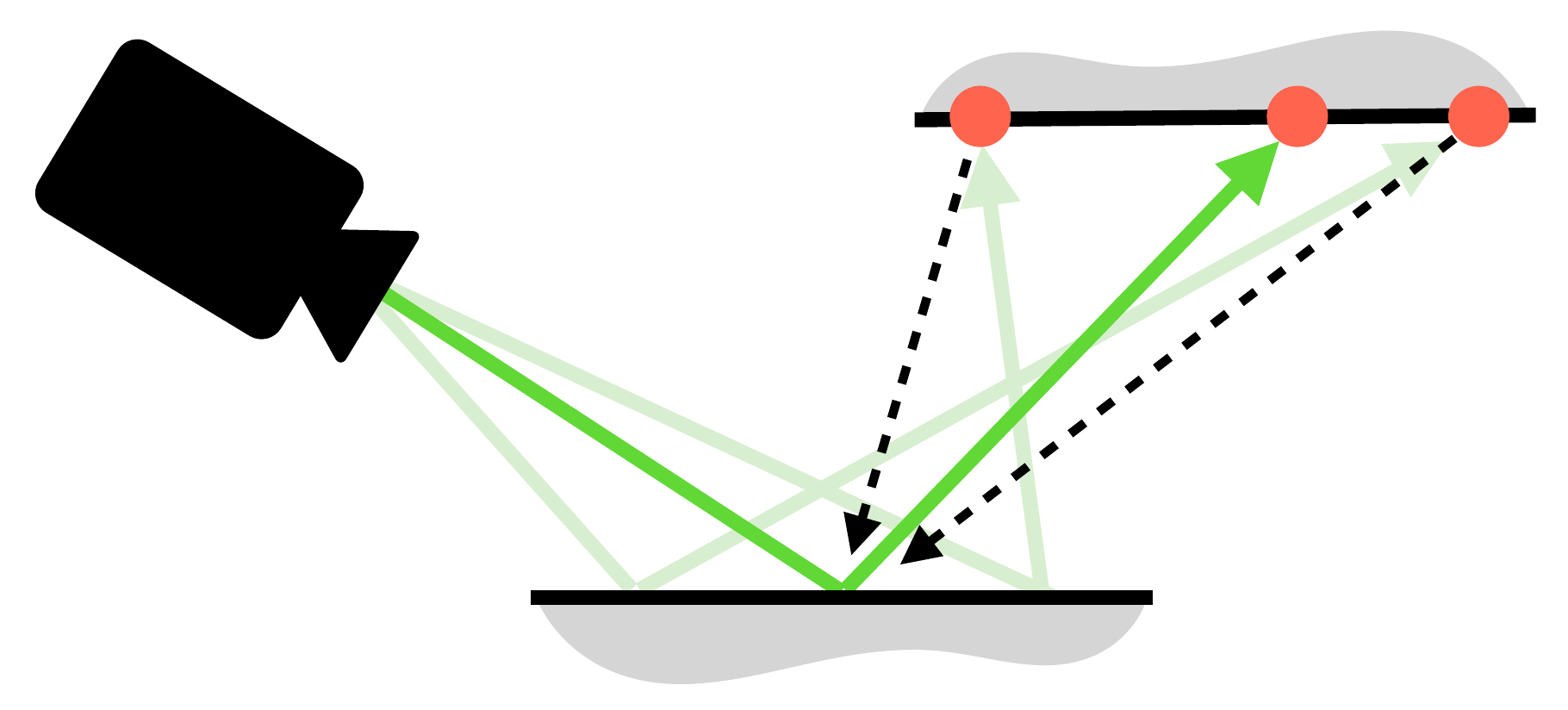}%
    \setlength{\unitlength}{0.011\linewidth}%
    \begin{picture}(0,0)
        \put(-63, 2) {$\pos_2$}
        \put(-48, 2) {$\pos_1$}
        \put(-33, 2) {$\pos_3$}
        \put(-10, 42) {$\posy_2$}
        \put(-20, 42) {$\posy_1$}
        \put(-40, 42) {$\posy_3$}
    \end{picture}\\[-6mm]
\end{wrapfigure}
In the illustration, we depict the simpler case of resampling among just 2 spatial candidates $\posy_2$ and $\posy_3$ for shading at the center vertex $\pos_1$.
For each candidate, we must query its contribution $\LddgiP(\posy_i, \posy_i\!\rightarrow\!\pos_1)$ to the center vertex, where ${\posy\!\rightarrow\!\pos}$ is the direction vector pointing from $\posy$ towards $\pos$ (dashed arrows).

These ${\text{\#vertices} \times \text{\#candidates}}$ queries can be avoided by approximating them with the $\text{\#vertices}$ values $\LddgiP(\posy_i, \posy_i\!\rightarrow\!\pos_i)$, which have to be evaluated for shading each vertex $\pos_i$ in any case.
This saves a factor of $\text{\#candidates}$ queries.
Importantly, using approximate values in the resampling step does not introduce additional bias in the rendered image -- merely additional noise.

Nonetheless, we analyze the error of this approximation by considering its three sources:
\begin{enumerate}
    \item the scene and the DDGI volume may change each frame,
    \item the shading location $\pos$ is different, and
    \item the view-path bias (Sec.~\ref{Sec:SelfShadowBias}) depends on the path prefix
\end{enumerate}

each of which become small if we apply the assumptions of our method: local pixel neighborhoods, near-diffuse scattering, and short frame times.
The approximation error is thus reasonably small in practice (see Fig.~\ref{fig:caching}) and we use this optimization in all our results.

\section{Results}

\begin{figure*}
\setlength{\fboxrule}{10pt}%
\setlength{\insetvsep}{20pt}%
\setlength{\tabcolsep}{-0.1pt}%
\renewcommand{\arraystretch}{1}%
\newlength\pinkroomoffset
\settowidth{\pinkroomoffset}{\PinkRoom}
\newlength\greekvillaoffset
\settowidth{\greekvillaoffset}{\GreekVilla}
\newlength\roomdooroffset
\settowidth{\roomdooroffset}{\RoomDoor}
\newlength\splitroomoffset
\settowidth{\splitroomoffset}{\SplitRoom}
\newlength\redballoffset
\settowidth{\redballoffset}{\RedBall}
\newlength\livingroomoffset
\settowidth{\livingroomoffset}{\LivingRoom}
\newlength\mapeheight
\settoheight{\mapeheight}{MAPE}
\footnotesize%
\hspace*{0.5mm}\begin{tabular}{cccccccc}
    & & \cite{Majercik2021ScalingGI} & \cite{bitterli20spatiotemporal} & \multicolumn{3}{c}{Ours} & \\
    \cmidrule(lr){3-3}
    \cmidrule(lr){4-4}
    \cmidrule(lr){5-7}
    &Reference
    &Primary DDGI
    &PT + ReSTIR
    &Secondary DDGI
    &+DDGI in ReSTIR
    &+Denoising
    &Reference
    \\%
    \setInset{A}{red}{679}{635}{225}{150}%
    \setInset{B}{orange}{1190}{625}{300}{200}%
    \hspace{-4mm}\rotatebox{90}{\hspace{-0.07875\textwidth}\hspace{-0.5\pinkroomoffset}\hspace{2.0\mapeheight}\PinkRoom}%
    &\addBeautyCrop{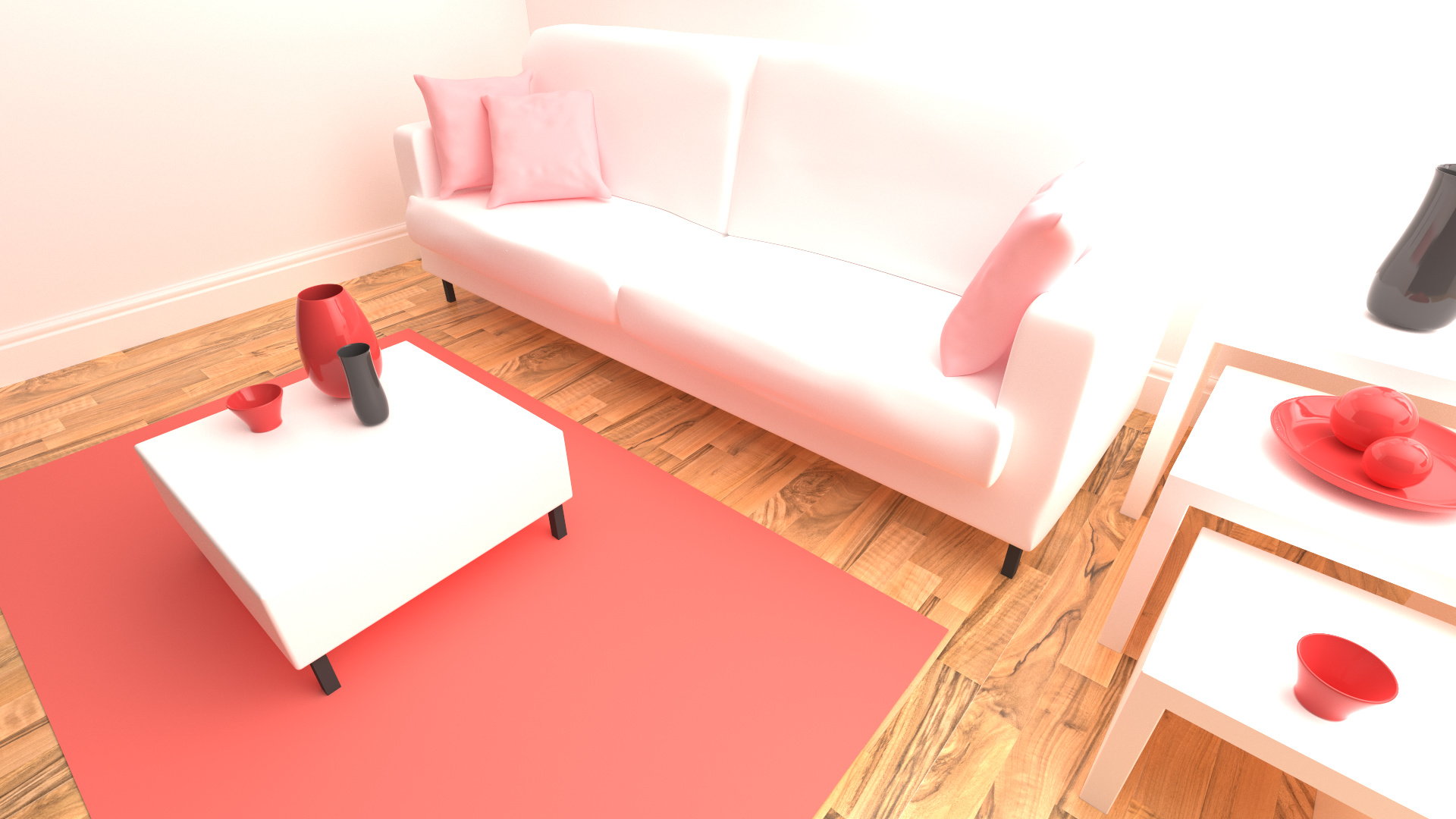}{0.28}{1920}{1080}{0}{0}{1920}{1080}%
    &\addInsets{generated_figures/fig_main_result_1/pink_room/ddgi.jpg}%
    &\addInsets{generated_figures/fig_main_result_1/pink_room/pt.jpg}%
    &\addInsets{generated_figures/fig_main_result_1/pink_room/ddgi-indirect.jpg}%
    &\addInsets{generated_figures/fig_main_result_1/pink_room/ddgi-indirect-resampled-bounded.jpg}%
    &\addInsets{generated_figures/fig_main_result_1/pink_room/ddgi-indirect-resampled-bounded-denoised.jpg}%
    &\addInsets{generated_figures/fig_main_result_1/pink_room/ref.jpg}%
    \\%
    & MAPE / Frametime (ms):
    &0.44 / 10.2 ms%
    &1.35 / 49.1 ms%
    &0.79 / 10.8 ms%
    &0.45 / 15.8 ms%
    &0.22 / 24.0 ms%
    &%
    \\%
    \setInset{A}{red}{640}{115}{150}{100}%
    \setInset{B}{orange}{574}{360}{150}{100}%
    \hspace{-4mm}\rotatebox{90}{\hspace{-0.07875\textwidth}\hspace{-0.5\greekvillaoffset}\hspace{2.0\mapeheight}\GreekVilla}%
    &\addBeautyCrop{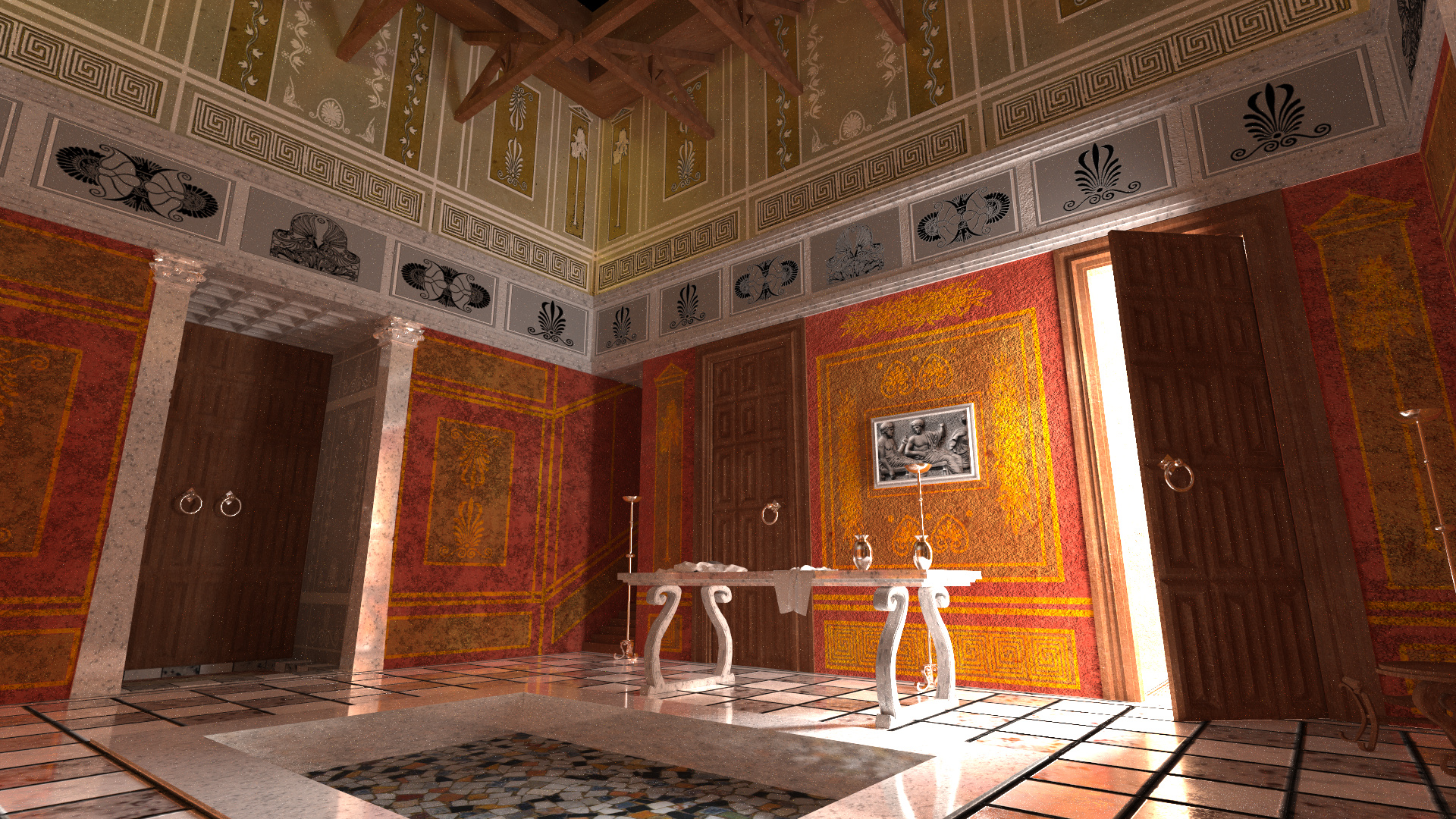}{0.28}{1920}{1080}{0}{0}{1920}{1080}%
    &\addInsets{generated_figures/fig_main_result_1/greek_villa/ddgi.jpg}%
    &\addInsets{generated_figures/fig_main_result_1/greek_villa/pt.jpg}%
    &\addInsets{generated_figures/fig_main_result_1/greek_villa/ddgi-indirect.jpg}%
    &\addInsets{generated_figures/fig_main_result_1/greek_villa/ddgi-indirect-resampled-bounded.jpg}%
    &\addInsets{generated_figures/fig_main_result_1/greek_villa/ddgi-indirect-resampled-bounded-denoised.jpg}%
    &\addInsets{generated_figures/fig_main_result_1/greek_villa/ref.jpg}%
    \\%
    & MAPE / Frametime (ms):
    &0.80 / 12.2 ms%
    &1.10 / 22.5 ms%
    &0.78 / 12.8 ms%
    &0.64 / 18.4 ms%
    &0.38 / 26.6 ms%
    &%
    \\%
    \setInset{A}{red}{1032}{382}{225}{150}%
    \setInset{B}{orange}{389}{177}{300}{200}%
    \hspace{-4mm}\rotatebox{90}{\hspace{-0.07875\textwidth}\hspace{-0.5\roomdooroffset}\hspace{2.0\mapeheight}\RoomDoor}%
    &\addBeautyCrop{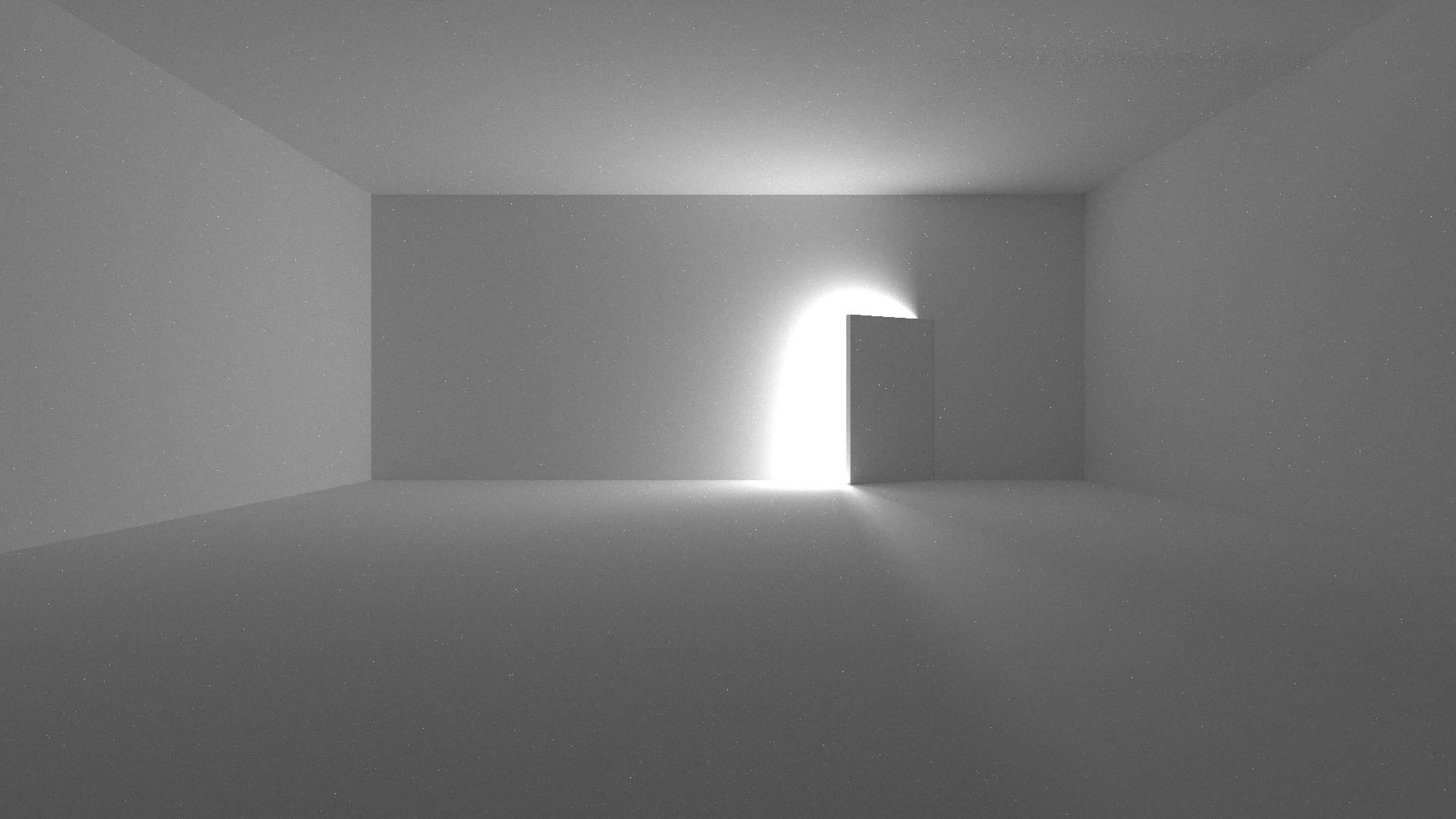}{0.28}{1920}{1080}{0}{0}{1920}{1080}%
    &\addInsets{generated_figures/fig_main_result_1/room_door/ddgi.jpg}%
    &\addInsets{generated_figures/fig_main_result_1/room_door/pt.jpg}%
    &\addInsets{generated_figures/fig_main_result_1/room_door/ddgi-indirect.jpg}%
    &\addInsets{generated_figures/fig_main_result_1/room_door/ddgi-indirect-resampled-bounded.jpg}%
    &\addInsets{generated_figures/fig_main_result_1/room_door/ddgi-indirect-resampled-bounded-denoised.jpg}%
    &\addInsets{generated_figures/fig_main_result_1/room_door/ref.jpg}%
    \\%
    & MAPE / Frametime (ms):
    &0.07 / 8.62 ms%
    &1.73 / 18.9 ms%
    &0.52 / 9.05 ms%
    &0.23 / 13.5 ms%
    &0.08 / 22.0 ms%
    &%
    \\%
    \setInset{A}{red}{350}{780}{300}{200}%
    \setInset{B}{orange}{750}{800}{300}{200}%
    \hspace{-4mm}\rotatebox{90}{\hspace{-0.07875\textwidth}\hspace{-0.5\splitroomoffset}\hspace{2.0\mapeheight}\SplitRoom}%
    &\addBeautyCrop{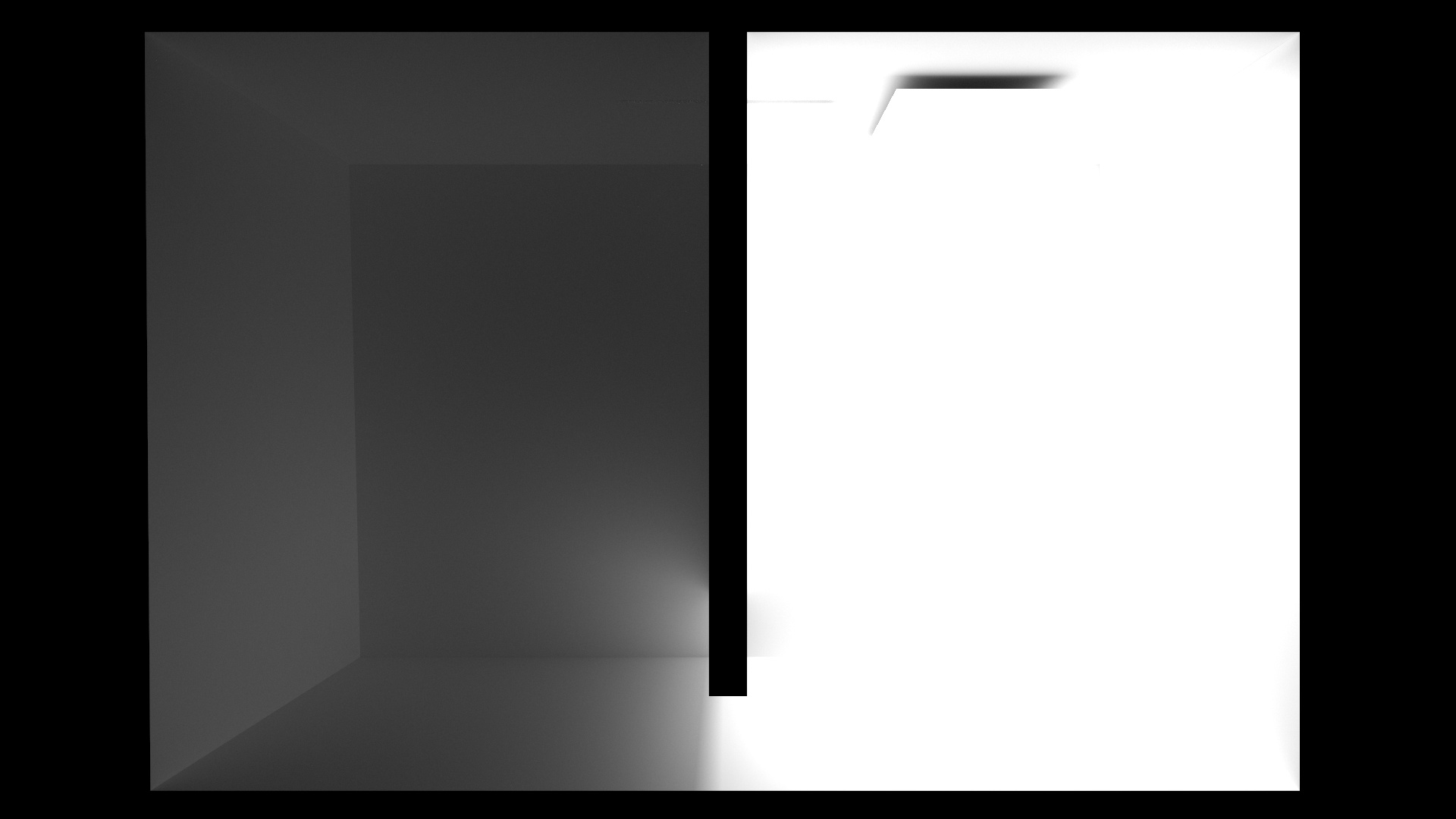}{0.28}{1920}{1080}{0}{0}{1920}{1080}%
    &\addInsets{generated_figures/fig_main_result_1/occlusion_resampling/ddgi.jpg}%
    &\addInsets{generated_figures/fig_main_result_1/occlusion_resampling/pt.jpg}%
    &\addInsets{generated_figures/fig_main_result_1/occlusion_resampling/ddgi-indirect.jpg}%
    &\addInsets{generated_figures/fig_main_result_1/occlusion_resampling/ddgi-indirect-resampled-bounded.jpg}%
    &\addInsets{generated_figures/fig_main_result_1/occlusion_resampling/ddgi-indirect-resampled-bounded-denoised.jpg}%
    &\addInsets{generated_figures/fig_main_result_1/occlusion_resampling/ref.jpg}%
    \\%
    & MAPE / Frametime (ms):
    &0.22 / 6.73 ms%
    &0.76 / 12.4 ms%
    &0.38 / 7.00 ms%
    &0.18 / 10.9 ms%
    &0.12 / 18.3 ms%
    &%
    \\%
    \setInset{A}{red}{950}{190}{300}{200}%
    \setInset{B}{orange}{1286}{713}{300}{200}%
    \hspace{-4mm}\rotatebox{90}{\hspace{-0.07875\textwidth}\hspace{-0.5\redballoffset}\hspace{2.0\mapeheight}\RedBall}%
    &\addBeautyCrop{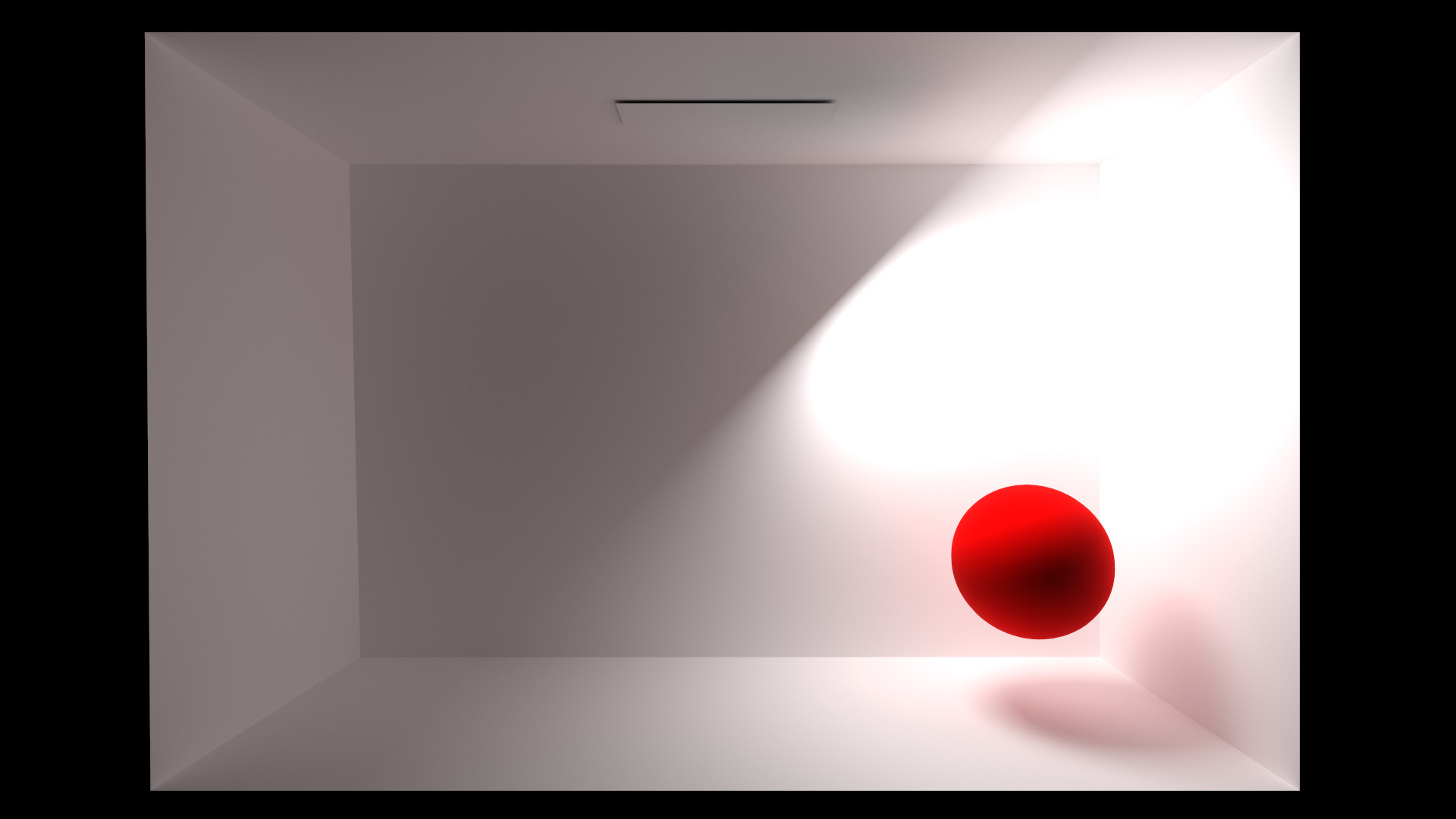}{0.28}{1920}{1080}{0}{0}{1920}{1080}%
    &\addInsets{generated_figures/fig_main_result_1/simple_resampling/ddgi.jpg}%
    &\addInsets{generated_figures/fig_main_result_1/simple_resampling/pt.jpg}%
    &\addInsets{generated_figures/fig_main_result_1/simple_resampling/ddgi-indirect.jpg}%
    &\addInsets{generated_figures/fig_main_result_1/simple_resampling/ddgi-indirect-resampled-bounded.jpg}%
    &\addInsets{generated_figures/fig_main_result_1/simple_resampling/ddgi-indirect-resampled-bounded-denoised.jpg}%
    &\addInsets{generated_figures/fig_main_result_1/simple_resampling/ref.jpg}%
    \\%
    & MAPE / Frametime (ms):
    &0.18 / 7.89 ms%
    &0.88 / 12.7 ms%
    &0.54 / 8.15 ms%
    &0.17 / 13.1 ms%
    &0.05 / 21.2 ms%
    &%
    \\%
    \setInset{A}{red}{100}{270}{225}{150}%
    \setInset{B}{orange}{1247}{309}{300}{200}%
    \hspace{-4mm}\rotatebox{90}{\hspace{-0.07875\textwidth}\hspace{-0.5\livingroomoffset}\hspace{2.0\mapeheight}\LivingRoom}%
    &\addBeautyCrop{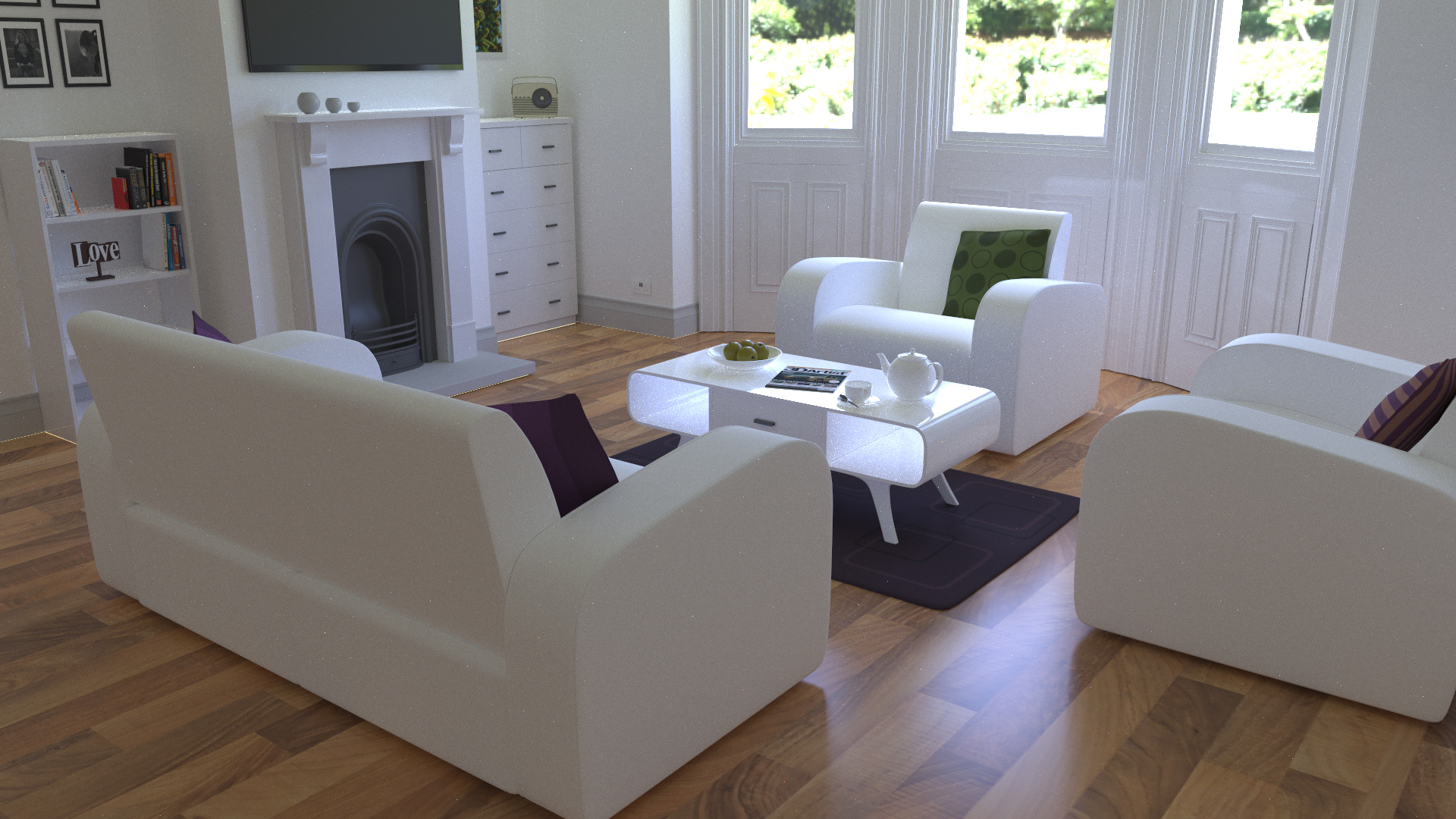}{0.28}{1920}{1080}{0}{0}{1920}{1080}%
    &\addInsets{generated_figures/fig_main_result_1/living_room/ddgi.jpg}%
    &\addInsets{generated_figures/fig_main_result_1/living_room/pt.jpg}%
    &\addInsets{generated_figures/fig_main_result_1/living_room/ddgi-indirect.jpg}%
    &\addInsets{generated_figures/fig_main_result_1/living_room/ddgi-indirect-resampled-bounded.jpg}%
    &\addInsets{generated_figures/fig_main_result_1/living_room/ddgi-indirect-resampled-bounded-denoised.jpg}%
    &\addInsets{generated_figures/fig_main_result_1/living_room/ref.jpg}%
    \\%
    & MAPE / Frametime (ms):
    &0.27 / 10.0 ms%
    &1.18 / 54.5 ms%
    &0.45 / 10.8 ms%
    &0.46 / 18.4 ms%
    &0.19 / 26.3 ms%
    &%
    \\%
\end{tabular}
    \caption{Comparison of rendering algorithms at 1 sample per pixel with respect to noise and bias.
    All algorithms are based on an unbiased path tracer that uses ReSTIR~\cite{bitterli20spatiotemporal} to estimate direct illumination at the primary vertex.
    As compared to this baseline, dynamic diffuse global illumination (Primary DDGI)
    renders smooth indirect illumination. Bias artifacts inherent with the DDGI
    approximation can be ameliorated by pushing the query of DDGI one bounce further into the light path (Secondary DDGI).
    Facilitating the ReSTIR sampling mechanism to jointly importance sample these DDGI queries with direct illumination, the induced noise is reduced (+DDGI in ReSTIR), and is further amenable to commodity denoising methods (+Denoising).
    We also show a failure case (\LivingRoom{}) where the outer walls consist of flat planes that receive strong sunlight from the outside.
    In this difficult scene, our method is unable to hide DDGI light leaking artifacts (red inset) and BSDF importance sampling (Secondary DDGI) performs slightly better than ReSTIR sampling.}
    \label{fig:main-result1spp}
\end{figure*}

We implemented DDGI resampling in Direct3D 12 using the Falcor rendering framework~\cite{Benty20}.
All results were generated on a high-end desktop machine with an Intel i7-6800K CPU and an NVIDIA RTX 3090 GPU.

Fig.~\ref{fig:main-result1spp} summarizes our results, comparing visuals, render time, and mean absolute percentage error (MAPE) of 5 different rendering techniques at 1 sample per pixel.
MAPE is defined as $\frac{1}{N} \sum_{i=1}^N |v_i - \hat{v}_i| / (\hat{v}_i + \epsilon)$, where $\hat{v}_i$ is the value of the $i$-th pixel in the reference image, $v_i$ is the value of the $i$-th rendered pixel, and $\epsilon = 0.01$ prevents near-black pixels from dominating the metric.

All rendering techniques are based on an unbiased path tracer that uses ReSTIR~\cite{bitterli20spatiotemporal} to estimate direct illumination at the primary vertex.
\begin{enumerate}
    \item \textbf{PT + ReSTIR:} The aforementioned baseline path tracer.
    \item \textbf{Primary DDGI:} The original DDGI algorithm~\cite{Majercik2019Irradiance,Majercik2021ScalingGI}, which approximates \emph{indirect} illumination at the \emph{primary} vertex of the path.
    \item \textbf{Secondary DDGI:} Our augmented DDGI algorithm approximates \emph{global} illumination at the \emph{secondary} vertex of the path (Sec.~\ref{Sec:Augmentation}--\ref{Sec:SelfShadowBias}). Near-specular transport is traced recursively.
    \item \textbf{+ DDGI in ReSTIR:} The same as before, but the DDGI approximation is included in the ReSTIR algorithm (Sec.~\ref{Sec:Sampling}). The diffuse weak singularity is separately sampled (Sec.~\ref{Sec:Singularity}).
    \item \textbf{+ Denoising:} The same as before, but with OptiX denoising.
\end{enumerate}
As expected, the DDGI algorithm~\cite{Majercik2019Irradiance,Majercik2021ScalingGI} has the least noise with the greatest performance (``Primary DDGI'' column).
However, it also exhibits significant visual artifacts such as flat shading (e.g.,\ in the \GreekVilla{}), a lack of ambient occlusion (e.g.,\ in the \PinkRoom{}), and light leaking (e.g.,\ in the \LivingRoom{}).
At the other extreme, an unbiased path tracer produces highly accurate results but requires a long time for its noise to converge away in globally illuminated scenes, even when using ReSTIR~\cite{bitterli20spatiotemporal} for direct lighting (``PT + ReSTIR'' column).

\begin{table}
    \caption{\label{tab:equal_quality}
        Time to converge to equal MAPE
    }%
    \vspace{-2mm}
\setlength{\tabcolsep}{4pt}%
\renewcommand{\arraystretch}{0.75}%
\small%
\begin{tabularx}{\columnwidth}{llrrr}%
    \toprule
    Scene & Method &Frames&Time&MAPE\\
    \midrule
    \multirow{3}{*}{\GreekVilla{}}
    & PT+ReSTIR &129&2908.00 ms&0.573\\
    & +Secondary DDGI &8&113.52 ms&0.578\\
    & +DDGI Resampling &1&19.83 ms&0.572\\
    \\[-1.5mm]
    \multirow{3}{*}{\LivingRoom{}}
    & PT+ReSTIR &315&17157.04 ms&0.366\\
    & +Secondary DDGI &1&10.82 ms&0.366\\
    & +DDGI Resampling &3&56.57 ms&0.365\\
    \\[-1.5mm]
    \multirow{3}{*}{\PinkRoom{}}
    & PT+ReSTIR &29&1423.70 ms&0.384\\
    & +Secondary DDGI &9&97.21 ms&0.390\\
    & +DDGI Resampling &1&16.85 ms&0.383\\
    \\[-1.5mm]
    \multirow{3}{*}{\RedBall{}}
    & PT+ReSTIR &109&1381.71 ms&0.140\\
    & +Secondary DDGI &29&254.17 ms&0.142\\
    & +DDGI Resampling &1&13.37 ms&0.140\\
    \\[-1.5mm]
    \multirow{3}{*}{\RoomDoor{}}
    & PT+ReSTIR &4095&77278.05 ms&0.206\\
    & +Secondary DDGI &79&748.51 ms&0.200\\
    & +DDGI Resampling &1&13.28 ms&0.199\\
    \\[-1.5mm]
    \multirow{3}{*}{\SplitRoom{}}
    & PT+ReSTIR &150&1854.92 ms&0.162\\
    & +Secondary DDGI &27&202.42 ms&0.162\\
    & +DDGI Resampling &1&10.92 ms&0.161\\
    \midrule
    \multirow{3}{*}{Average}
        & PT+ReSTIR &804.50&17000.57 ms&0.305\\
        & +Secondary DDGI &25.50&237.77 ms&0.306\\
        & +DDGI Resampling &1.33&21.81 ms&0.303\\
    \bottomrule
\end{tabularx}
\end{table}

Our proposed \emph{indirect} use of DDGI probes -- provided that the probes contain the energy of the full transport (Sec.~\ref{Sec:Augmentation} and \ref{Sec:Glossy}) -- reduces noise while introducing only little bias, because the inaccuracies of the DDGI probes are hidden behind the first scattering event (``Secondary DDGI'' column).
Additionally, the performance is improved, because paths are terminated early into DDGI probes.
Incorporating the indirect lighting injected by the DDGI probes in the resampling step further reduces noise (``DDGI in ReSTIR'' column).
Overall, the noise is now low enough that adding the OptiX denoiser results in more visually pleasing images than the original DDGI algorithm, albeit at an on average 1.5-2$\times$ greater cost.

We also show a failure case (\LivingRoom{}) where the outer walls consist of flat planes that receive strong sunlight from the outside.
In this difficult scene, secondary DDGI queries are unable to hide DDGI's light leaking artifacts (red inset) and BSDF importance sampling (``Secondary DDGI'' column) performs slightly better than ReSTIR sampling.
This shows that further work is needed in making DDGI more resilient to thin geometry as well as in understanding the circumstances in which ReSTIR can be outperformed by its candidate generation strategies.

\paragraph*{Quantitative evaluation.}
To gauge the convergence improvements offered by using DDGI at the secondary path vertex (with or without resampling), we list the time it takes for all noisy methods to reach equal error in Tab.~\ref{tab:equal_quality}.
In all scenes, except for the \LivingRoom{}, DDGI resampling converges the fastest -- in some scenes by a significant margin -- even when taking into account the slightly increased cost of rendering.

To better understand this cost, we break it down by component in Tab.~\ref{tab:timings}.
As expected, terminating paths into DDGI reduces the ray tracing cost because paths are shorter, even when taking into account the additional cost of querying DDGI during the shading step, as well as the cost of updating the DDGI volume every frame.
However, perhaps unexpectedly, resampling only becomes marginally more expensive by incorporating DDGI in it (by $1.7$ ms on average) thanks to the caching of previous DDGI queries (Sec.~\ref{sec:caching}).
The remaining overhead of DDGI resampling arises in tracing \& shading of (near-)specular path suffixes, which must now be performed twice per pixel: once for the BSDF sample that estimates $\lfloor T_{\bsdf} \rfloor$ and once for the resampled vertex that estimates $\lceil T_{\bsdf} \rceil$ (see Sec.~\ref{Sec:Singularity}).

\begin{table}
    \caption{\label{tab:timings}
     Rendering cost by component
    }%
    \vspace{-2mm}
\setlength{\tabcolsep}{4pt}%
\renewcommand{\arraystretch}{0.75}%
\small%
\begin{tabularx}{\columnwidth}{lllll}%
    \toprule
    Scene & Method 
        &
        \hspace{-6mm}
                Trace \& shade
        &
            \hspace{-3mm}
            Resampling
        &
                \hspace{-2mm}
        Update
    \\
    \midrule
    \multirow{3}{*}{\GreekVilla{}}
    & PT+ReSTIR &\hspace{-1mm}20.38 ms&\hspace{-1mm}1.10 ms&\hspace{-1mm}---\\
    & +Secondary DDGI &\hspace{-1mm}10.65 ms&\hspace{-1mm}1.20 ms&\hspace{-1mm}1.54 ms\\
    & +DDGI Resampling &\hspace{-1mm}14.03 ms&\hspace{-1mm}3.13 ms&\hspace{-1mm}1.55 ms\\
    \\[-1.5mm]
    \multirow{3}{*}{\LivingRoom{}}
    & PT+ReSTIR &\hspace{-1mm}52.72 ms&\hspace{-1mm}1.42 ms&\hspace{-1mm}---\\
    & +Secondary DDGI &\hspace{-1mm}8.18 ms&\hspace{-1mm}1.62 ms&\hspace{-1mm}0.57 ms\\
    & +DDGI Resampling &\hspace{-1mm}13.32 ms&\hspace{-1mm}4.47 ms&\hspace{-1mm}0.57 ms\\
    \\[-1.5mm]
    \multirow{3}{*}{\PinkRoom{}}
    & PT+ReSTIR &\hspace{-1mm}46.60 ms&\hspace{-1mm}1.08 ms&\hspace{-1mm}---\\
    & +Secondary DDGI &\hspace{-1mm}8.08 ms&\hspace{-1mm}1.26 ms&\hspace{-1mm}1.13 ms\\
    & +DDGI Resampling &\hspace{-1mm}11.99 ms&\hspace{-1mm}2.41 ms&\hspace{-1mm}1.14 ms\\
    \\[-1.5mm]
    \multirow{3}{*}{\RedBall{}}
    & PT+ReSTIR &\hspace{-1mm}10.71 ms&\hspace{-1mm}0.95 ms&\hspace{-1mm}---\\
    & +Secondary DDGI &\hspace{-1mm}5.47 ms&\hspace{-1mm}1.27 ms&\hspace{-1mm}1.02 ms\\
    & +DDGI Resampling &\hspace{-1mm}8.57 ms&\hspace{-1mm}3.23 ms&\hspace{-1mm}1.01 ms\\
    \\[-1.5mm]
    \multirow{3}{*}{\RoomDoor{}}
    & PT+ReSTIR &\hspace{-1mm}16.91 ms&\hspace{-1mm}1.10 ms&\hspace{-1mm}---\\
    & +Secondary DDGI &\hspace{-1mm}6.71 ms&\hspace{-1mm}1.50 ms&\hspace{-1mm}1.07 ms\\
    & +DDGI Resampling &\hspace{-1mm}9.64 ms&\hspace{-1mm}2.33 ms&\hspace{-1mm}1.08 ms\\
    \\[-1.5mm]
    \multirow{3}{*}{\SplitRoom{}}
    & PT+ReSTIR &\hspace{-1mm}10.90 ms&\hspace{-1mm}0.88 ms&\hspace{-1mm}---\\
    & +Secondary DDGI &\hspace{-1mm}5.33 ms&\hspace{-1mm}1.24 ms&\hspace{-1mm}0.44 ms\\
    & +DDGI Resampling &\hspace{-1mm}7.63 ms&\hspace{-1mm}2.65 ms&\hspace{-1mm}0.44 ms\\
    \midrule
    \multirow{3}{*}{Average}
        & PT+ReSTIR &\hspace{-1mm}26.37 ms&\hspace{-1mm}1.09 ms&\hspace{-1mm}---\\
        & +Secondary DDGI &\hspace{-1mm}7.40 ms&\hspace{-1mm}1.35 ms&\hspace{-1mm}0.96 ms\\
        & +DDGI Resampling &\hspace{-1mm}10.86 ms&\hspace{-1mm}3.04 ms&\hspace{-1mm}0.97 ms\\
    \bottomrule
\end{tabularx}
\end{table}

\section{Discussion and Future Work} \label{Sec:Discussion}

The following discussion focuses on future avenues for potential performance and noise improvements.

\paragraph*{Subsampling.}
To reduce the remaining resampling cost, it might be feasible to subsample the generation of new candidates: for example, only every $n$-th pixel could generate novel samples from the BSDF and from light sources.
All other pixels would merely resample among spatio-temporal neighbors.
Since this strategy is rather orthogonal to our work -- it would benefit all ReSTIR-type algorithms -- we did not include it in our technique.
However, we feel that it warrants a detailed analysis in the future.

\paragraph*{Alternative Caching Strategies.}
While we focused on the specific use of DDGI, we believe other caching strategies, such as the recent neural radiance caching~\cite{mueller2021realtime}, can similarly benefit from importance sampling through ReSTIR.\@
Another example is the concurrent work of Ouyang et al.~\cite{ReSTIRGI} who apply ReSTIR to a cache that is effectively a collection of spatio-directional point light sources.
Their approach represents a different point on the Pareto front that trades off noise, bias, and performance -- one that has lower bias (and can even be unbiased) at the cost of tracing longer paths, leading to extra noise.

\subsection{Alternative Variance Reduction Strategies}

In real-time rendering, sampling with low variance is key due to the limited budget of rays to be traced.
In this context, we also analyzed the efficiency of utilizing DDGI as a control variate, as well as that of using it for path guiding instead of resampling.
Both approaches were overall less efficient than the presented DDGI resampling, but we will nonetheless describe them in the hope that future algorithms may build on top of them.

\begin{figure}
\setlength{\fboxrule}{1pt}%
\setlength{\tabcolsep}{1pt}%
\renewcommand{\arraystretch}{1}%
\small%
\centering%
\begin{tabular}{cccccccc}%

                   \hspace{-2cm} &
                    
                   \hspace{-2cm}DDGI Resampling &
                    
                   \hspace{-2cm} &
                    
                    \hspace{-2.25cm}+ Control Variate
                \\%

    \multicolumn{4}{c}{\setlength{\fboxsep}{1.5pt}\begin{overpic}[width=\linewidth]{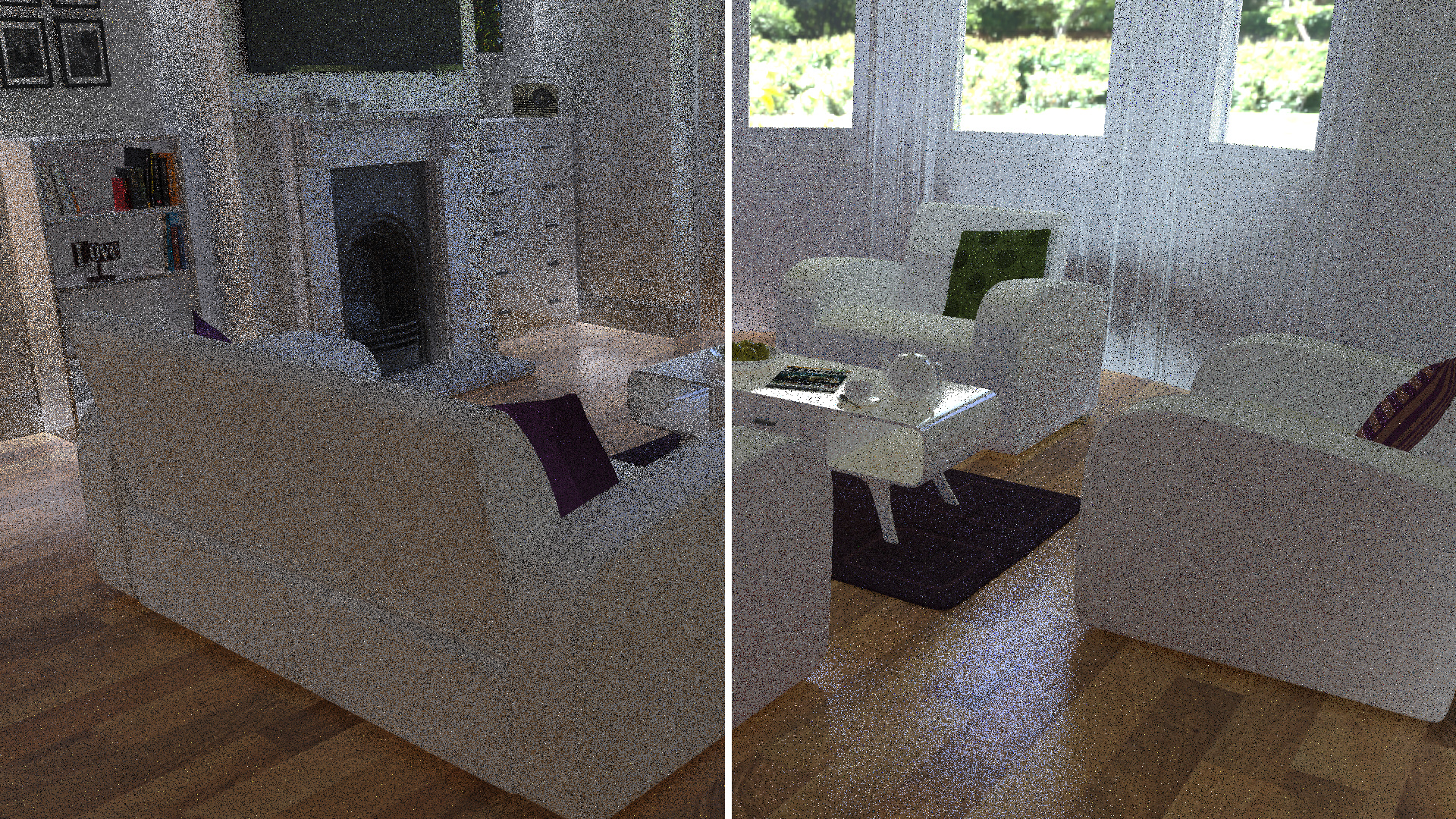}
        \put(78.38541666666666, 31.510416666666664){\makebox(0,0){\tikz\draw[orange,ultra thick] (0,0) rectangle (0.10312500000000001\linewidth, 0.041249999999999995\linewidth);}}
        \put(31.614583333333336, 32.86458333333333){\makebox(0,0){\tikz\draw[red,ultra thick] (0,0) rectangle (0.10312500000000001\linewidth, 0.041249999999999995\linewidth);}}
    \end{overpic}}
    \\%
    \fcolorbox{orange}{orange}{\includegraphics[width=0.235\linewidth]{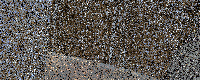}} &
    \fcolorbox{orange}{orange}{\includegraphics[width=0.235\linewidth]{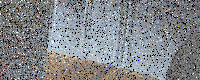}} &
    \fcolorbox{orange}{orange}{\includegraphics[width=0.235\linewidth]{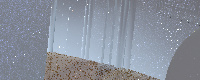}} &
    \fcolorbox{orange}{orange}{\includegraphics[width=0.235\linewidth]{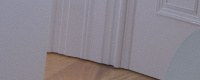}}
    \\%
    \fcolorbox{red}{red}{\includegraphics[width=0.235\linewidth]{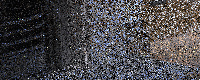}} &
    \fcolorbox{red}{red}{\includegraphics[width=0.235\linewidth]{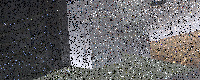}} &
    \fcolorbox{red}{red}{\includegraphics[width=0.235\linewidth]{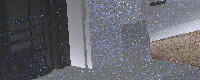}} &
    \fcolorbox{red}{red}{\includegraphics[width=0.235\linewidth]{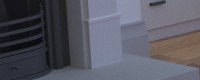}}
    \\%
                        DDGI Resampling &
                                + Control Variate &
                                Visualization &
                                Reference\\ &
                                & of DDGI &
                \\%
\end{tabular}
    \caption{Analysis of DDGI Resampling (left split) augmented by DDGI used as a control variate at the primary vertex (right split). 
    The transient artifacts visible at low sampling rates are due to the underlying probe volume as can be seen by the visualization of DDGI.
    At 1 spp, these are too salient to ignore, although with higher sample counts both DDGI Resampling and DDGI Resampling with control variate converge to the same result.}
    \label{fig:CVcomparison}
\end{figure}

\paragraph*{DDGI as Control Variate.}
For scenes with simple occlusion, the original DDGI data structure provides a low-bias, noise-free result. This can sometimes be desired over our new resampling, which will introduce noise even in cases where the resulting lower bias is not visually noticeable. To combat this,
we explored extending our solution by using primary hitpoint irradiance $\EddgiP$ as a control variate of diffuse transport.
That is, we assumed that ${\frac{\rho}{\pi}\,\EddgiP}$ at primary vertices yields the correct indirect illumination and we resampled proportional to the remaining non-diffuse transport \emph{plus} (neglecting constants for brevity) the absolute difference between ${(\LddgiP + L_e)}$ at the secondary vertex and the spatio-directionally nearest DDGI update ray among those that were traced while rendering the last frame.
This amounts to using the DDGI update rays as a spatio-directionally piecewise constant control variate and their average, $\EddgiP$, as its exact integral.

When shading, we then applied DDGI to the primary vertex (as described in \cite{Majercik2019Irradiance}) and added {\em the difference} (not the absolute value) between ${L_e + \LddgiP}$ at the secondary vertex and the nearest DDGI update ray.
Adding this difference in the shading step may result in negative radiance values, which frequently manifest as negative fireflies that cause black pixels.
We omitted spatio-temporal probe interpolation for
the DDGI control variate, as otherwise the value of $\LddgiP$ at the primary vertex
would not be an exact integral of the control variate.

Fig.~\ref{fig:CVcomparison} shows our results.
Though this algorithm converges to the same (biased) solution as plain DDGI resampling, at one sample per pixel, the 
control variate exposes light leaks similar to the na\"\i{}ve
ReSTIR + DDGI combination. At the same time, the noise level does not improve upon plain DDGI resampling, which produces less prominent
light leaks.

While using DDGI as a control variate may be beneficial
in some settings,
in general, it appears not sufficiently efficient and remains an avenue for future research.

\paragraph*{DDGI for Path Guiding.}
Based on the DDGI data structure, guiding rays towards where
the radiance comes from is straightforward. Given a position
in space, we can query the closest DDGI probe and build the
cumulative distribution function (CDF) from the results of
its update rays to sample a guiding direction.
While the results of the probe rays are stored, building
the CDF and sampling from it was too slow
to increase the overall efficiency of our algorithm.

In addition, using $\LddgiP$ to normalize the PDF and sample the guiding direction 
in an unbiased way requires the same modifications to be made to the DDGI volume that were 
made to the control variate. These make the volume less stable and thus less effective for path guiding.
An alternative would be porting efficient path guiding data structures~\cite{Vorba2019PGP}
to the GPU.

\paragraph*{Combining Primary and Secondary DDGI.}
Even though we focus on hiding the bias of DDGI behind scattering interactions, there are situations in which DDGI's bias is very low in the first place and does not warrant the extra noise caused by secondary queries; see the orange inset of the \RedBall scene in Fig.~\ref{fig:main-result1spp}.
It likely pays off to design an automatic mechanism for detecting such situations and to select among primary and secondary DDGI queries accordingly.

\paragraph*{Path Space Filtering.}
Lastly, alternatively to screen-space denoising, we believe that additional spatio-temporal reuse through path space filtering could be beneficial.
While ReSTIR amounts to spatio-temporal filtering of sampling probabilities and DDGI amounts to spatio-temporal filtering of radiance at \emph{secondary} vertices, path space filtering would provide additional spatio-temporal filtering of radiance at \emph{primary} vertices.
Recent hash-based implementations~\cite{binder2021massively} can run with little overhead on modern GPUs.

\subsection{Limitations}
The limitations of our approach are largely inherited from its components: ReSTIR and DDGI.\@
ReSTIR is a screen-space technique whose reuse capability depends highly on the availability of motion vectors, slow camera motion, and transparency.
These downsides can be mitigated by transitioning to a world-space representation~\cite{regir}, but another limitation remains: through resampling, samples become spatio-temporally correlated, which reduces the total amount of information available to modern denoisers.
Effective denoising in the presence of ReSTIR samples is still ongoing work.

The limitations induced by DDGI are two-fold.
First, being grid-based, DDGI does not scale easily to expansive scenes where the camera may only focus on small regions.
And second, DDGI is incapable of resolving fine spatio-directional detail such as intricate caustics, shadows, ambient occlusion or near-specular reflections.
These limitations are hidden to a degree when DDGI is queried after a primary scattering interaction that is sufficiently diffuse or after undergoing sufficiently many secondary interactions, but better quality can be achieved by using a more accurate cache~\cite{mueller2021realtime}.

\section{Conclusion}

We have combined DDGI~\cite{Majercik2021ScalingGI} and ReSTIR~\cite{bitterli20spatiotemporal}, which, at first glance, might seem orthogonal.
DDGI renders indirect illumination and ReSTIR direct illumination.
However, the appearance is deceiving and a small modification to the DDGI algorithm -- querying the probe volume at secondary path vertices rather than primary ones -- unifies the two approaches into an algorithm that performs much better than its constituent parts.
By querying DDGI at secondary vertices, it acts as a scene-spanning light source, effectively turning the global illumination problem into a much simpler direct illumination problem, which ReSTIR can holistically attack.
This observation is a general one that has inspired rendering algorithms since the 90s, making us excited about future work in this direction.

Quantitatively, the combined algorithm is about 60\% more expensive than either of the two prior works, but makes up for its cost in terms of low noise and bias.
It achieves equal error over an order of magnitude more quickly.
Together with denoising, real-time rendering of global illumination in dynamic scenes is feasible.

\section*{Acknowledgements}
We would like to thank David Luebke and Aaron Lefohn for supporting this work. We also thank Benedikt Bitterli and Chris Wyman for providing the initial codebase in Falcor~\cite{Benty20}, Chris Wyman and Peter Shirley for many early ideas and fruitful discussion, and Nikolaus Binder and Christoph Schied for proofreading the article and many valuable comments. The Pink Room scene was modified from ``The Modern Living Room" by Wig42 and the Living Room scene was modified from ``The White Room" by Jay-Artist. The Greek Villa was modified from ``3D Greek Villa" by holy\_diver.


\printbibliography                


\end{document}